\documentclass{aa}  

\usepackage{graphicx}
\usepackage{txfonts}
\usepackage{epsfig}
\usepackage{xspace}
\usepackage{multicol}
\usepackage{multirow}

\newcommand{\Nu}{\textit{NuSTAR}\xspace}
\newcommand{\XMM}{\textit{XMM-Newton}\xspace}
\newcommand{\Ch}{\textit{Chandra}\xspace}
\newcommand{\flux}{\,erg\,s$^{-1}$\,cm$^{-2}$\xspace} % cgs flux
\newcommand{\kms}{\,km\,s$^{-1}$} % kilometres per second
\newcommand{\Msunyr}{$\mathrm{M}_\odot$\,yr$^{-1}$\xspace} % Solar mass per year
\newcommand{\Rsun}{$\mathrm{R}_\odot$\xspace} % Solar radius

\begin{document} 

    \title{Evidence for non-thermal X-ray emission from the double WR colliding-wind binary \textit{Apep}}
    
    \author{S. del Palacio\inst{1}
            \and
            F. Garc\'ia\inst{2}
            \and 
            M. De Becker\inst{3} 
            \and 
            D. Altamirano\inst{4} 
            \and 
            V. Bosch-Ramon\inst{5}
            \and 
            \\
            P. Benaglia\inst{2}  
            \and 
            B. Marcote\inst{6} 
            \and 
            G.~E. Romero\inst{2}
            }
    
    \institute{Department of Space, Earth and Environment, Chalmers University of Technology, SE-412 96 Gothenburg, Sweden\\
    \email{santiago.delpalacio@chalmers.se}
    \and
    Instituto Argentino de Radioastronom\'ia (CCT La Plata, CONICET; CICPBA; UNLP), C.C.5, (1894) Villa Elisa, Buenos Aires, Argentina.
    \and
   Space sciences, Technologies and Astrophysics Research unit -- STAR, University of Li\`ege, Quartier Agora, 19c, All\'ee du 6 Ao\^ut, B5c, B-4000 Sart Tilman, Belgium
    \and
    School of Physics and Astronomy, University of Southampton Highfield Campus, Southampton SO17 1PS, UK.
    \and
    Departament de F\'{i}sica Qu\`antica i Astrof\'{i}sica, Institut de Ci\`encies del Cosmos (ICC), Universitat de Barcelona (IEEC-UB), Mart\'{i} i Franqu\`es 1, E08028 Barcelona, Spain.
    \and
    Joint Institute for VLBI ERIC, Oude Hoogeveensedijk 4, 7991 PD Dwingeloo, the Netherlands.
                }

   \date{Received ---; accepted ---}

% \abstract{}{}{}{}{} 
% 5 {} token are mandatory
 
  \abstract
   {Massive colliding-wind binaries (CWBs) can be non-thermal sources. The emission produced in their wind-collision region (WCR) encodes information of both the shocks properties and the relativistic electrons accelerated in them. The recently discovered system \textit{Apep}, a unique massive system hosting two Wolf-Rayet stars, is the most powerful synchrotron radio emitter among the known CWBs, being an exciting candidate to investigate the non-thermal processes associated with stellar wind shocks.}
   {We intend to break the degeneracy between the relativistic particle population and the magnetic field strength in the WCR of \textit{Apep} by probing its hard X-ray spectrum, where inverse-Compton (IC) emission is expected to dominate.}
   {We observe \textit{Apep} with \Nu for 60~ks and combine this with a re-analysis of a deep archival \XMM observation to better constrain the X-ray spectrum. We use a non-thermal emission model to derive physical parameters from the results.}
   {We detect hard X-ray emission consistent with a power-law component from \textit{Apep}. This is compatible with IC emission produced in the WCR for a magnetic field of $\approx 105$--190~mG, corresponding to a magnetic-to-thermal pressure ratio in the shocks of $\approx 0.007$--0.021, and a fraction of $\sim 1.5\times10^{-4}$ of the total wind kinetic power being transferred to relativistic electrons.} 
   {This is the first time that the non-thermal emission from a CWB is detected both in radio and high energies. This allows us to derive the most robust constraints of the particle acceleration efficiency and magnetic field intensity in a CWB so far, reducing the typical uncertainty of a few orders of magnitude to just within a factor of a few. This constitutes an important step forward in our characterisation of the physical properties of CWBs.} 


   \maketitle
%
%-------------------------------------------------------------------

%===================================================================
\section{Introduction}
%===================================================================

Colliding-wind binaries (CWBs) are binary systems in which the powerful winds of the massive stars collide. The strong shocks at the wind-collision region (WCR) produce very hot ($>10^6$~K) X-ray emitting plasma. Morever, they can also accelerate relativistic particles \citep{Eichler1993, Benaglia2003} and constitute a subset of objects called Particle-Accelerating Colliding-Wind Binaries \citep[PACWBs;][]{DeBecker2013}. The efficiency of this particle acceleration process is however still poorly constrained both theoretically and observationally. The usually assumed scenario for particle acceleration in PACWBs is Diffusive Shock Acceleration \citep[DSA;][]{Drury1983}. 

Relativistic electrons, expected in general to radiate more efficiently than relativistic protons, can up-scatter stellar optical/ultraviolet photons to X-ray or $\gamma$-ray emission by the Inverse Compton (IC) process. Relativistic electrons can also radiate synchrotron emission in the radio band by interacting with the magnetic fields in the WCR. 
Many CWBs present non-thermal radio emission \citep{DeBecker2013}, but this is insufficient to characterise both the relativistic electron population and the magnetic field intensity in the emitter without severe partitioning assumptions \citep{DeBecker2018}. Meanwhile, detections at hard X-rays and above remain scarce: $\eta$-Car has been clearly detected in both hard X-rays \citep{Hamaguchi2018} and $\gamma$-rays \citep{Tavani2009, Reitberger2015, HESS2020, Marti-Devesa2021}, while $\gamma^2$~Vel has been recently confirmed as a $\gamma$-ray source \citep{Marti-Devesa2020}, and a tentative detection of non-thermal hard X-rays ($E \lesssim 18$~keV) has been associated with HD~93129A \citep{delPalacio2020}. 

The X-ray spectral energy distribution (SED) of a CWB is determined by the thermal and the non-thermal radiation components, which depend on the WCR properties, together with the local wind absorption \citep{Pittard2010}. The emission from individual stellar winds can only produce soft X-rays at energies $\lesssim 1$~keV, and the total absorption from most stellar winds and the interstellar medium (ISM) is not relevant above 2~keV. Thus, the SED at energies $> 3$~keV is determined solely by processes in the WCR. Thermal processes are likely to dominate the SED up to $\sim10$~keV given the high wind velocities and consequent post-shock temperatures, and thus the non-thermal processes can only be investigated at energies above 10~keV. It is therefore necessary to have a broadband measurement of the X-ray SED to disentangle these two components. 

The system \textit{Apep} is a peculiar case of a massive binary made up of two Wolf-Rayet stars \citep{Callingham2019}. The stars are separated by more than 100~AU, which allows the stellar winds to accelerate to full speed before collision.
Radio observations of this system revealed that it is a very powerful synchrotron source \citep{Callingham2019}, which also establishes it as an efficient particle accelerator. In addition, \cite{Marcote2021} confirmed that this emission raises from the WCR using very long baseline interferometric observations. Further constraints on the radio spectrum by \cite{Bloot2022} allowed \cite{delPalacio2022} to model the source broadband emission in order to infer properties of the stellar winds and predict the SED of the source at high energies. However, these predictions are highly degenerate as it is not possible to disentangle the relativistic particle energy distribution and the magnetic field strength in the WCR, $B_\mathrm{WCR}$, solely from radio data \citep{delPalacio2022}. A recent analysis of \textit{Fermi}-LAT data in $\gamma$-rays placed stronger constraints on the high-energy SED of \textit{Apep}, but were still unable to detect its emission \citep{Marti-Devesa2022}.

In addition, \textit{Apep} was also observed in soft X-rays on different occasions with \textit{XMM-Newton} and \textit{Chandra}. \cite{Callingham2019} analysed this data-set and concluded that this source: i) is point-like in X-rays; ii) is not variable on scales of years; and iii) has a predominantly thermal spectrum with a significant absorption below 2~keV. 

Here we aim to investigate the hard X-ray spectrum of \textit{Apep} and search for signatures of a non-thermal IC component, as predicted by \cite{delPalacio2022}. Measuring such a component can better constrain both the energy budget in relativistic particles and the magnetic field strength in the WCR. 
With this purpose we conducted observations of \textit{Apep} with \Nu, probing for the first time its spectrum at energies $>10$~keV. In this work we present the analysis and interpretation of such observations.

%===================================================================
\section{Observations and data reduction}
%===================================================================

%-----------------------------------------------------
\subsection{The system Apep} 
%-----------------------------------------------------

\begin{table*}
    \caption{Parameters of the WC+WN system \textit{Apep}.}
    \centering
    \begin{tabular}{@{}lll@{}}
    \hline\hline
    Parameter 		            &   Value			            &	Reference		\\
    \hline%
    Distance			        &	$d=2.4^{+0.2}_{-0.5}$~kpc	&   \cite{Callingham2019} \\ 
    Projected system separation	&   $D_\mathrm{proj}=47\pm6$~mas&   \cite{Han2020} \\
    Projection angle            &   $\psi=85^\circ$             &   \cite{delPalacio2022} \\
    Wind momentum rate ratio	&	$\eta=0.44\pm0.08$	        &   \cite{Marcote2021}	\\
    \hline
    Stellar temperature         &	$T_\mathrm{eff,WN}=65\,000$~K &  Typical \citep[e.g.][]{Crowther2007, Hamann2019}	\\
    Stellar radius  		    &	$R_\mathrm{WN} = 6$~\Rsun		&     Typical \citep[e.g.][]{Hamann2019}\\    
    Wind terminal velocity      &	$v_{\infty,\mathrm{WN}} = 3500\pm100$~\kms 	&   \cite{Callingham2020} \\ 
    Wind mass-loss rate         &	$\dot{M}_\mathrm{WN} = (4\pm1)\times10^{-5}$~\Msunyr & \cite{delPalacio2022} \\ 
    Wind mean atomic weight     &	$\mu_\mathrm{WN} = 2.0$ 		        &	Typical \citep[e.g.][]{Leitherer1995}	\\ 
    %Wind temperature            &	$T_\mathrm{w,\mathrm{WN}} = 0.3\,T_\mathrm{eff,\mathrm{WN}}$ & Typical \citep[e.g.][]{Drew1990} \\ 
    %Wind filling factor         &	$f_\mathrm{WN} = 0.2$	                &   Typical \citep[e.g.][]{Runacres2002}	\\
    \hline
    Stellar temperature         &	$T_\mathrm{eff,\mathrm{WC}}=60\,000$~K & Typical \citep[e.g.][]{Crowther2007, Sander2019}	\\
    Stellar radius  		    &	$R_\mathrm{WC} = 6.3$~\Rsun		&      Typical \citep[e.g.][]{Sander2019}            \\    
    Wind terminal velocity      &	$v_{\infty,\mathrm{WC}} = 2100\pm200$~\kms 	&	\cite{Callingham2020}		\\ 
    Wind mass-loss rate         &	$\dot{M}_\mathrm{WC} = (2.9\pm0.7) \times10^{-5}$~\Msunyr & \cite{delPalacio2022}	\\ 
    Wind mean atomic weight     &	$\mu_\mathrm{WC} = 4.0$ 		        &	Typical \citep[e.g.][]{Cappa2004}\\ 
    %Wind temperature            &	$T_\mathrm{w,\mathrm{WC}} = 0.3\,T_\mathrm{eff,\mathrm{WC}}$ 	&	Typical \citep[e.g.][]{Drew1990} \\ 
    %Wind volume filling factor         &	$f_\mathrm{WC} = 0.2$	                &   Typical \citep[e.g.][]{Runacres2002}	\\ 
    \hline\hline
    \end{tabular}
    \label{tab:system_parameters}
\end{table*}

The massive binary \textit{Apep} (2XMM~J160050.7$-$514245) is located at $RA = 10^\mathrm{h}\,43^\mathrm{m}\, 57.5^\mathrm{s}$, $DEC=-59\degr 32\arcmin 51.4\arcsec$ (J2000). Its orbit is wide, with a separation between the stars of tens of AU \citep{Han2020}. The primary star is a WN star while the secondary is a WC star. These stars have very massive and fast winds with kinetic powers of $L_\mathrm{WN}\approx1.5\times10^{38}$~erg\,s$^{-1}$ and $L_\mathrm{WC}\approx4.1\times10^{37}$~erg\,s$^{-1}$ (with an uncertainty of $\approx 30$\%; see Table~\ref{tab:system_parameters}). This constitutes an abundant energy reservoir to feed emission processes at the WCR. In fact, the WCR in \textit{Apep} is exceptionally luminous, being the brightest PACWB detected at radio wavelengths \citep{Callingham2019}. A more detailed list of the relevant system parameters is given in Table~\ref{tab:system_parameters}. 

This source has been observed in X-rays with both \XMM and \Ch during 2015--2021. Most of these observations have been previously analysed by \cite{Callingham2019}, who showed that the source does not present significant variability in X-rays. We observed this system for the first time with \Nu in 2022. In addition, we complemented this with a re-analysis of a deep \XMM observation in order to characterise better the X-ray SED of \textit{Apep}. We summarise the analysed observations in Table~\ref{tab:observations} and describe them in more detail in the following subsections.

\setlength{\tabcolsep}{3.0pt}
\begin{table*}
\caption{Summary of the X-ray observations analysed. In the case of \Nu, A and B refer to FPMA and FPMB, respectively.}
\label{tab:observations}
\begin{tabular}{l c c c c c c c c}
\hline\hline\\[-7pt]
Instrument      &   Obs.~ID     & Date (start)    & Exposure time (ks)  &  Effective time (ks) & Offset (')\\[2pt] %
\hline
\Nu &   30402001002 & 2022-06-17  &     31.0    & 31.0 (A), 30.8 (B) & 1.36 (A), 1.99 (B) \\[2pt] %
\Nu &   30402001004 & 2022-06-18  &     30.1    & 29.3 (A), 28.8 (B) &  1.43 (A), 2.06 (B) \\[2pt] %
%\hline
\XMM &  0742050101   & 2015-03-08 &   105 (PN), 137 (MOS2)     &  79.9 (PN), 106.3 (MOS2) & 8.8  \\
\hline
\end{tabular}
\end{table*}

%-----------------------------------------------------
\subsection{\XMM} \label{sec:xmm}
%-----------------------------------------------------

\textit{Apep} is in the field of view of several archival \XMM observations. Of these, Obs. 0742050101 is the deepest one ($> 100$~ks) and therefore the one we chose to analyse. The major drawback of this observation is the large offset from on-axis to the position of \textit{Apep}, which is $8.8\arcmin$. The source appears in both PN and MOS2 cameras, and the observation was carried out in full frame mode. Other details of this observation are summarised in Table~\ref{tab:observations}.

Data processing was performed using the Science Analysis Software
\texttt{SAS v.20.0.0} and the calibration files (CCF) available in August 2022. We used the metatasks \texttt{emproc} and \texttt{epproc} to reduce the data. We then filtered periods of high background or soft proton flares. Standard screening criteria were adopted, namely pattern $\leq 12$ for MOS and pattern $\leq 4$ for PN. We determined good time intervals by selecting events with \texttt{PI$>$10000} and \texttt{PATTERN$==$0} and adopting the standard rejection thresholds \texttt{RATE$\leq$0.35} for MOS2 and \texttt{RATE$\leq$0.4} for PN. The effective time after filtering is reported in Table~\ref{tab:observations}. These values are 25--30~ks shorter than those of \cite{Callingham2019}, which suggests that our selection of GTIs was more conservative. 

To produce the spectra, the radius of the extraction region was set to 50\arcsec, as the source is $8.8\arcmin$ off axis and therefore the PSF is larger than for on-axis sources (for which typically 10--30\arcsec\, is used). 
The background spectrum was extracted in an elliptical region located on the same chip, in an area devoid of point sources, selected using the \texttt{ebkreg} task. However, the adopted background region has a negligible impact on the results given that the source is very bright. Adequate response matrix files (RMF) and ancillary response files (ARF) were produced using the dedicated tasks ({\texttt rmfgen} and {\texttt arfgen}, respectively) for all spectra. On this last point, we note that the standard psfmodel \texttt{ELLBETA} does not work well for MOS2 due to \textit{Apep} being a very bright source and quite off-axis ($8.8\arcmin$), so we used the \texttt{psfmodel=EXTENDED} option (\textit{XMM} support, priv. comm.). This correction improves the match between the MOS2 and the PN spectra (a mismatch around 6~keV can be seen in Supplementary Information Fig.~4 from \citealt{Callingham2019}, where the standard ELLBETA model was used for MOS2). Other parameters adopted for the ARF file were \texttt{extendedsource=no}, \texttt{detmaptype=psf}, and \texttt{applyabsfluxcorr=yes}, being the last one used to improve the cross-calibration between \XMM and \Nu. We finally grouped the spectra using the task \texttt{ftgrouppha} with \texttt{grouptype=opt}.

%-----------------------------------------------------
\subsection{\Nu} \label{sec:nustar}
%-----------------------------------------------------

The \textit{NuSTAR} X-ray observatory was launched in 2012 and its major asset is its unique imaging capacity in hard X-rays. The observatory includes two co-aligned X-ray grazing incidence telescopes, known as FPMA and FPMB for their focal plane modules, which are comprised of four rectangular solid state CdZnTe detectors. \textit{NuSTAR} is capable of observing in the 3--79 keV energy range with an angular resolution of 18\arcsec \citep[half power diameter of 58\arcsec;][]{NuSTAR2013}.

We observed the massive binary \textit{Apep} in June 2022 with \Nu under program 8020 (PI: del Palacio). The observations were carried out in two 30-ks visits, adding up to roughly 60~ks of exposure time with both cameras. We refer to the observations in each epoch as 2022a and 2022b. We summarise the relevant details of these observations in Table~\ref{tab:observations}.

We reduced the data using \texttt{Heasoft 6.30.1} and the latest calibration files available in June 2022 (CALDB 4.9.7-0). 
We used the \texttt{nupipeline} task to create level 2 data products with the options \texttt{saacalc=2}, \texttt{saamode=optimized} and \texttt{tentacle=yes} to filter high background epochs. This led to negligible data loss in the 2022a observations and $< 3\%$ data loss in the 2022b observation\footnote{\url{http://www.srl.caltech.edu/NuSTAR_Public/NuSTAROperationSite/SAA_Filtering/SAA_Filter.php}.}. We then used the \texttt{nuproducts} task to create level 3 data products. We extracted the source spectrum from a 55\arcsec\, region centred in \textit{Apep}, while the background was extracted from an ellipse located in the same chip, sufficiently far from the source as to avoid contamination. The selected background region also avoids contamination from the supernova remnant G330.2+1.0, as shown in Fig.~\ref{fig:xmm_image}.
Further analysis of the influence of the selected background region is presented in Appendix~\ref{appendix:background}.
Finally, we binned the spectra using the task \texttt{ftgrouppha} with the option \texttt{grouptype=opt}.

%-----------------------------------------------------
\subsection{\texttt{XSPEC} spectral model} \label{sec:spectral_model}
%-----------------------------------------------------

Once the spectra have been obtained, one needs to fit them with a spectral model in order to extract physical information. 
Any model we adopt should be both physically-motivated and simple enough as to reproduce the data without requiring a very large number of parameters. A spectral model for a PACWB should include an absorption component --that can take into account both internal absorption in the stellar winds and external absorption in the ISM--, a thermal component --dominated by the WCR emission--, and a non-thermal component --relevant only at $E > 10$~keV, also produced by the WCR--.
The thermal emission from CWBs is usually approximated using an \textit{apec} model \citep{Smith2001}, because it is simple and can reproduce well the emission from an optically thin plasma. Multiple \textit{apec} components can be used to emulate the temperature gradient along the WCR \citep[e.g.][]{Pittard2010}. However, the \textit{apec} model assumes that electrons and ions are in equilibrium, and the shocks in the WCR can be collisionless under certain conditions, leading to an ionisation that can be out-of-equilibrium. This depends on the relation between two timescales: the dynamical timescale, $t_\mathrm{dyn}$, and the electron-ion temperature equalisation timescale, $t_\mathrm{eq}$. 
The timescale $t_\mathrm{dyn}$ depends on the characteristic size of the WCR ($\sim D$) and on the velocity at which the material is advected away ($\sim v_\infty$), while $t_\mathrm{eq}$ depends on the post-shock temperature and density, and therefore on $v_\infty$ and $\dot{M}$. 
If $t_\mathrm{dyn} < t_\mathrm{eq}$, the electrons and ions cannot get in thermal equilibrium through Coulomb interactions before the post-shock plasma is advected away. This condition can be summarised through the parameter 
\mbox{$\zeta_\mathrm{eq} = \dfrac{t_\mathrm{dyn}}{t_\mathrm{eq}} \approx 
\dfrac{13.36}{\bar{\mu} \, \mu^{1.5}} 
\left( \dfrac{\dot{M}}{10^{-6}\,M_\odot}\right) 
\left(\dfrac{V_\infty}{1000\,\mathrm{km\,s^{-1}}}\right)^5 \left(\dfrac{10^{14}\,\mathrm{cm}}{D}\right)$},
such that if $\zeta_\mathrm{eq} < 1$ the difference between electron and ion temperatures should be taken into account \citep{Zhekov2000}. For long period binaries this condition is more likely to be fulfilled, such as in the case of \textit{Apep}. Actually, for the conditions in the shocks of the primary and secondary and using the parameters given in Table~\ref{tab:system_parameters}, we obtain $\zeta \sim 0.01 - 0.1$. Thus, the use of a non-equilibrium model, such as \textit{pshock}, is justified in this case and we explored this possibility as well.

Regarding the non-thermal emission, the simplest way to parameterise it is as a power-law component. We can constrain its spectral index considering that this component is expected to be IC radiation emitted by the same relativistic electron population in the WCR that is responsible for the synchrotron emission observed in the radio band \citep{delPalacio2022}. For a flux density $S_\nu \propto \nu^\alpha$ in the radio band, the specific photon flux density $F \propto E^{-\Gamma}$ in X-rays has a spectral index $\Gamma = -\alpha + 1$ \citep{delPalacio2022}. In the case of \textit{Apep}, a value of $\alpha = -0.72$ was reported by \cite{Callingham2019}, which leads to $\Gamma = 1.72$. We note that this value is slightly steeper than the canonical $\alpha = -0.5$ ($\Gamma = 1.5$) expected to arise from electrons accelerated by DSA in high Mach number shocks under the test-particle assumption. Nonetheless, this trend of steeper spectra is also seen in supernova remnants and can be related to the back-reaction of cosmic rays in the shocks \citep[e.g.][]{Drury1983, Gabici2019}. 

Finally, the emitted X-ray radiation can be absorbed intrinsically in the source and externally in the ISM. In \texttt{XSPEC}, the standard model used to calculate the ISM absorption is \textit{TBabs}. The value of the $N_\mathrm{H}$ column can be taken from \cite{HI4PI2016}\footnote{\url{https://heasarc.gsfc.nasa.gov/cgi-bin/Tools/w3nh/w3nh.pl}}. For \textit{Apep}, the value retrieved is $N_\mathrm{H}=1.63\times 10^{22}$~cm$^{-2}$. Additional intrinsic absorption (mostly by the stellar winds) can be included using a \textit{phabs} model. Throughout this work the confidence intervals are obtained using the \texttt{error} command in \texttt{XSPEC} and given at a 1-$\sigma$ level unless stated otherwise. 

%===================================================================
\section{Results} \label{sec:results}
%===================================================================

We now focus on the analysis and model fitting of the spectra obtained with the \Nu and \XMM observatories. 

%-----------------------------------------------------
\subsection{\XMM}
%-----------------------------------------------------

\begin{figure}
    \centering
    \includegraphics[width=\linewidth]{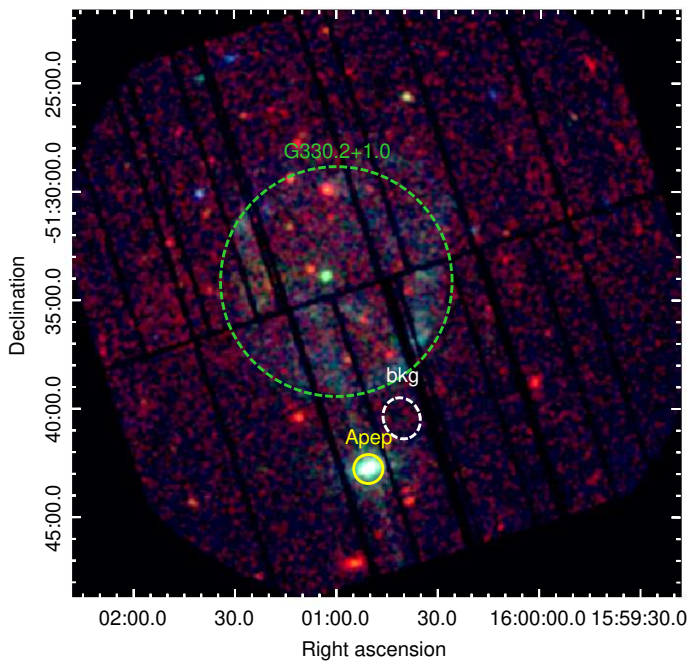}
    \caption{RGB \XMM image of the field of view of \textit{Apep} (red = 0.3--1.2 keV, green = 1.2--2.5 keV, blue = 2.5--8 keV) from the EPIC-PN detector. We also mark the position of the supernova remnant G330.2+1.0.}
    \label{fig:xmm_image}
\end{figure}

In Fig.~\ref{fig:xmm_image} we show an RGB exposure-corrected image of the field of view of \XMM. In the figure we also show the source and background extraction regions.

\begin{figure*}
    \centering
    \includegraphics[angle=0,width=0.49\linewidth]{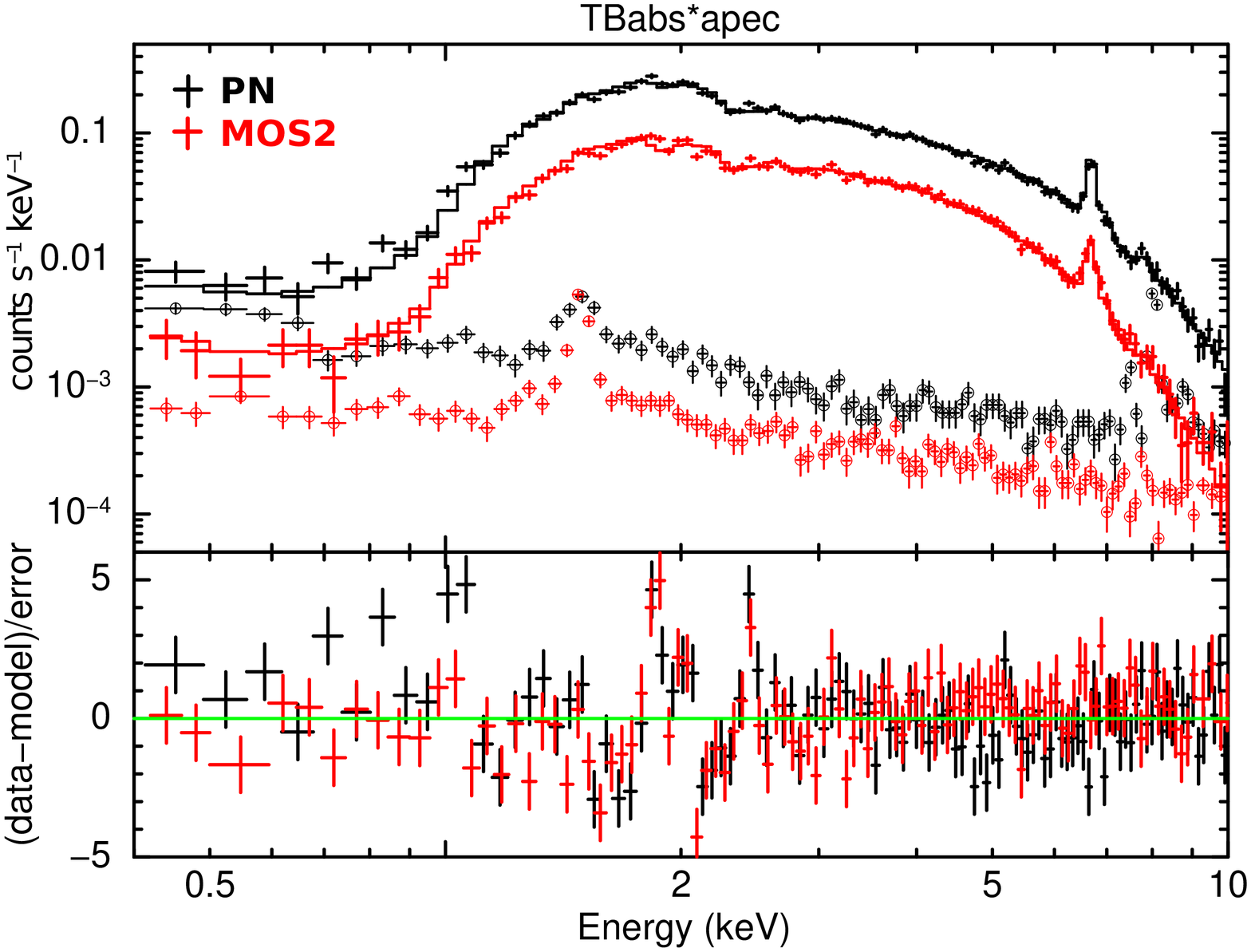}
    \includegraphics[angle=0,width=0.49\linewidth]{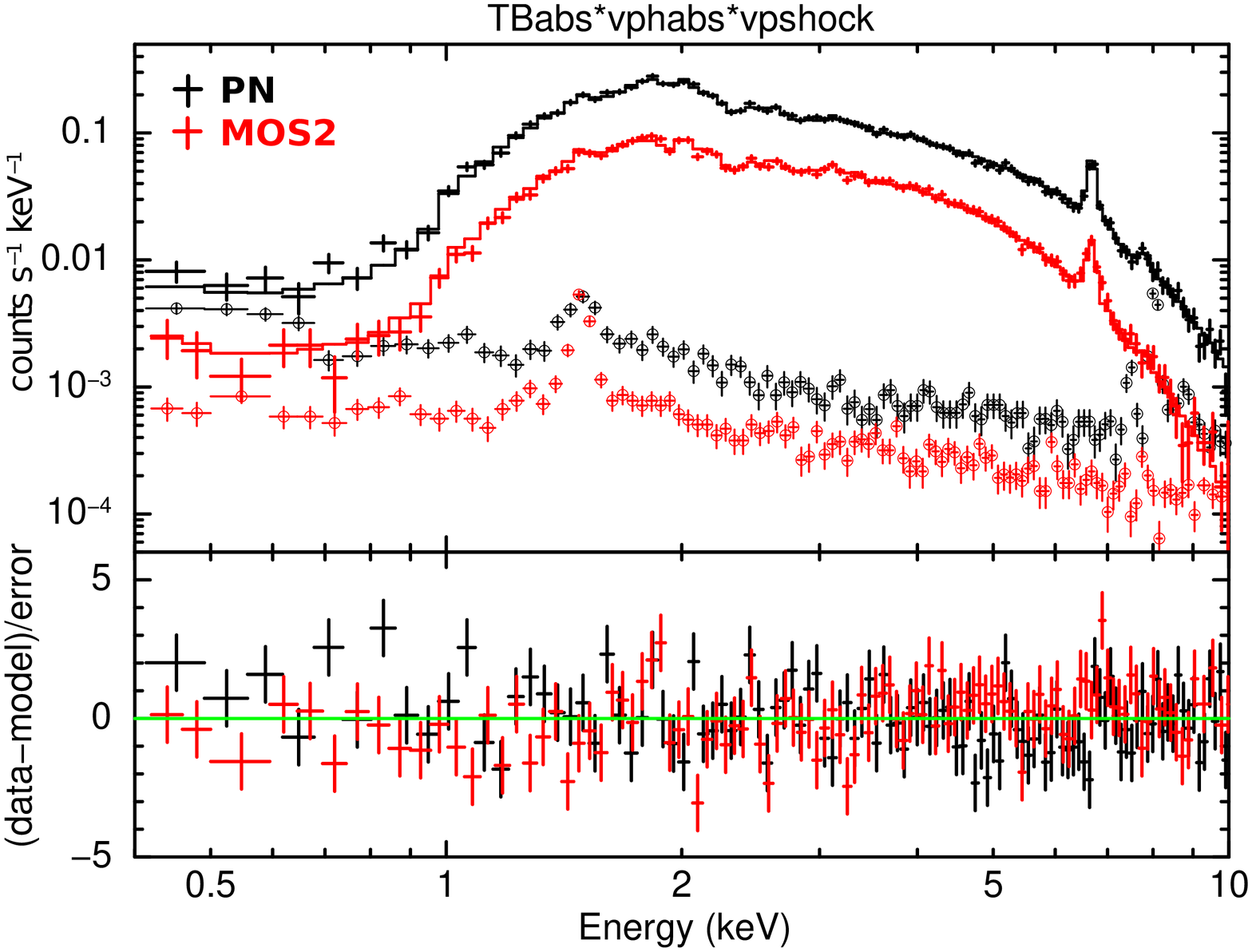}
    \caption{\textit{Apep} \XMM spectra in the 0.3--10~keV energy range. In the left panel the fitted model is \textit{TBabs*apec}, while in the right panel it is \textit{TBabs*vphabs*vpshock}. The latter model is preferred as it leads to smaller residuals between 1--2.5~keV.}
    \label{fig:xmm_spectrum_fit}
\end{figure*}

To fit the \XMM spectra, we first considered a model of the form \textit{constant*TBabs*apec}, with abundances set to \texttt{Wilms} \citep{Wilms2000} as done by \cite{Callingham2019}. The normalisation constant was set to unity for PN and then fitted to MOS2, obtaining $C = 1.06 \pm 0.01$. We show the fitted spectra in Fig.~\ref{fig:xmm_spectrum_fit}. In general, we found very similar results to \cite{Callingham2019} ($kT \approx 5.1$~keV, $N_{\rm H} \approx 2.7\times 10^{22}$~cm$^{-2}$, abundance $A \approx 0.5$, observed flux 
$F_\mathrm{0.3-10\, keV} \approx 8\times10^{-12}$~\flux), despite the differences in the data reduction (Sect.~\ref{sec:xmm}). Nonetheless, the goodness of the fit was actually quite poor, yielding large structured residuals around 1--2.5~keV at energies coincident with known Si and S transitions (Fig.~\ref{fig:xmm_spectrum_fit}) and a C-Stat 439.0/224. This motivated us to look for a better model.

As discussed in Sect.~\ref{sec:spectral_model}, adopting a \textit{pshock} model instead of an \textit{apec} is physically justified in the case of \textit{Apep}. The use of a \textit{pshock} model required only one additional free parameter, namely the upper limit on ionisation timescale ($\tau_\mathrm{u}$), and it improved significantly the quality of the fit (C-Stat 311.5/223). We note that this affected only the low energy portion of the spectrum, in particular by improving the ratios in the Si and S lines. This suggests that ionisation was indeed out of equilibrium, and is also consistent with the value obtained of $\tau_\mathrm{u} < 10^{12}$\,s\,cm$^3$. However, this had a completely negligible ($\sim 1\%$) impact on the spectra at energies above 3 keV. Further improvement can be done by setting variable abundances: although \cite{Callingham2019} claimed that using a \textit{vapec} model did not improve the fit significantly, we found that a \textit{vpshock} model can indeed introduce a significant improvement, reaching C-Stat 270.9/220. This can also be appreciated in Fig.~\ref{fig:xmm_spectrum_fit}, which highlights the smaller residuals retrieved in the 1--2.5~keV energy range with this model. When fitting the individual abundances, we obtained for Fe a similar value to that of $A$, indicating that this element dominates the value of $A$ for fixed relative abundances. A few elements presented different abundances (mainly Ne), while others like C and N were not constrained by the data and were left fixed to one. At last, as the stellar winds are expected to contribute significantly to the (photoelectric) absorption, we changed the absorption model to \textit{TBabs*vphabs}, fixing $N_\mathrm{H}=1.63\times10^{22}$\,cm$^{-2}$ for the \textit{TBabs} component (Sect.~\ref{sec:spectral_model}) and fixing the abundances of the \textit{vphabs} to those of the \textit{vpshock} model. This improved the fit slightly (C-Stat 268.4/220) without adding extra free parameters, and is therefore our preferred model for the \XMM data. In Table~\ref{tab:fit_parameters} we present in detail all the fitted parameters for the most relevant models, including the data from \Nu as discussed in the next section. Taking as a reference the PN camera, the observed flux in the 0.3--10~keV energy range is $F_\mathrm{0.3-10\,keV,obs}=(7.92\pm0.10)\times10^{-12}$~\flux, the ISM-unabsorbed flux is $F_\mathrm{0.3-10\,keV,unabs}=(9.69\pm0.05)\times10^{-12}$~\flux, and the unabsorbed flux from the \textit{vpshock} component only is $F_\mathrm{0.3-10\,keV,vpshock}=(1.89\pm0.04)\times10^{-11}$~\flux.

%-----------------------------------------------------
\subsection{\Nu} 
%-----------------------------------------------------

\begin{figure*}
    \centering
    \includegraphics[width=\linewidth]{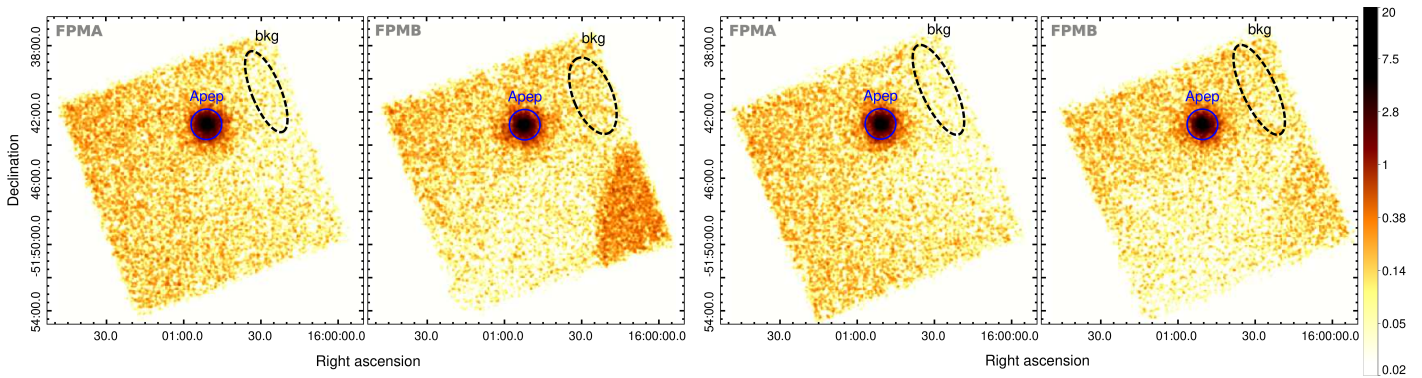}
    \caption{\Nu image in the 3--20~keV energy range for the 2022a (\textit{left panels}) and 2022b (\textit{right panels}) observations. For each epoch and camera we show the selected source and background extraction regions. Straylight can be seen at the bottom right corner of the FPMB observations.}
    \label{fig:nustar_image}
\end{figure*}

In Fig.~\ref{fig:nustar_image} we present an image in the 3--20 keV energy range ($PI$ channels 35--460) with \Nu for each observing epoch and camera. \textit{Apep} is clearly detected and no other bright sources appear in the field. We note the presence of straylight in the FPMB observations, although this is not problematic given that \textit{Apep} is far away from it. In Fig.~\ref{fig:nustar_image} we also show the selected source and background extraction regions. 

The obtained spectra for each epoch and camera are shown together in Fig.~\ref{fig:nustar_spectrum_compare} for comparison. All observations reveal a very similar spectrum in which the source is detected above the background up to $\gtrsim 20$~keV. We calculated the integrated source flux in the 3--10 keV and 10--25 keV energy ranges for each observation to make a quantitative comparison\footnote{For this test we adopted the models described in Sect.~\ref{sec:spectral_model} and used later in Sect.~\ref{sec:results}, and for all of them we reached the same conclusion.}. In all cases the fluxes differ in less than 10\% and are compatible within 1-$\sigma$ level.
We checked whether it was possible to combine the data from both observations to increase the signal-to-noise. For this we co-added the spectra of both cameras for each epoch using the \texttt{addspec} task with the options \texttt{qaddrmf=yes qsubback=yes} and the default value \texttt{bexpscale=1000}. We repeated the calculation of the integrated fluxes and obtained variations below 2\% ($< 1\sigma$ difference), indicating that they are perfectly compatible, which is to be expected considering that the observations were taken very close in time to each other.

We then co-added the spectra of both epochs for each camera to compare if there was any systematic difference between FPMA and FPMB. We obtained that the FPMA fluxes in the 3--10 keV were slightly ($\approx 5\%$) higher than for FPMB (with a 1-$\sigma$ significance), while in the 10--25 keV range the fluxes between both cameras match up to 2\% and this difference is less significant ($< 1\sigma$). 
This is within the calibration uncertainties of the instrument \citep{Madsen2015, Madsen2022}.
From these tests we concluded that both observations are compatible but with a small difference between the cameras. We therefore decided to work with both \Nu observations co-added for each camera separately to improve the signal-to-noise at the highest energies while preventing cross-calibration errors to introduce further errors to the fitting of the spectra. 
In Fig.~\ref{fig:nustar_spectrum_compare} we show the spectrum obtained for the combined observations. In this case the statistics are improved, as expected, and the source is brighter than the background up to $\sim$25~keV. In the fitting we include data up to 35~keV as there is valuable spectral information between 25--35~keV, and we make use of Cash statistics in \texttt{XSPEC} to deal properly with the low number of counts in this energy range. 

\begin{figure*}
    \centering
    \includegraphics[width=0.49\linewidth]{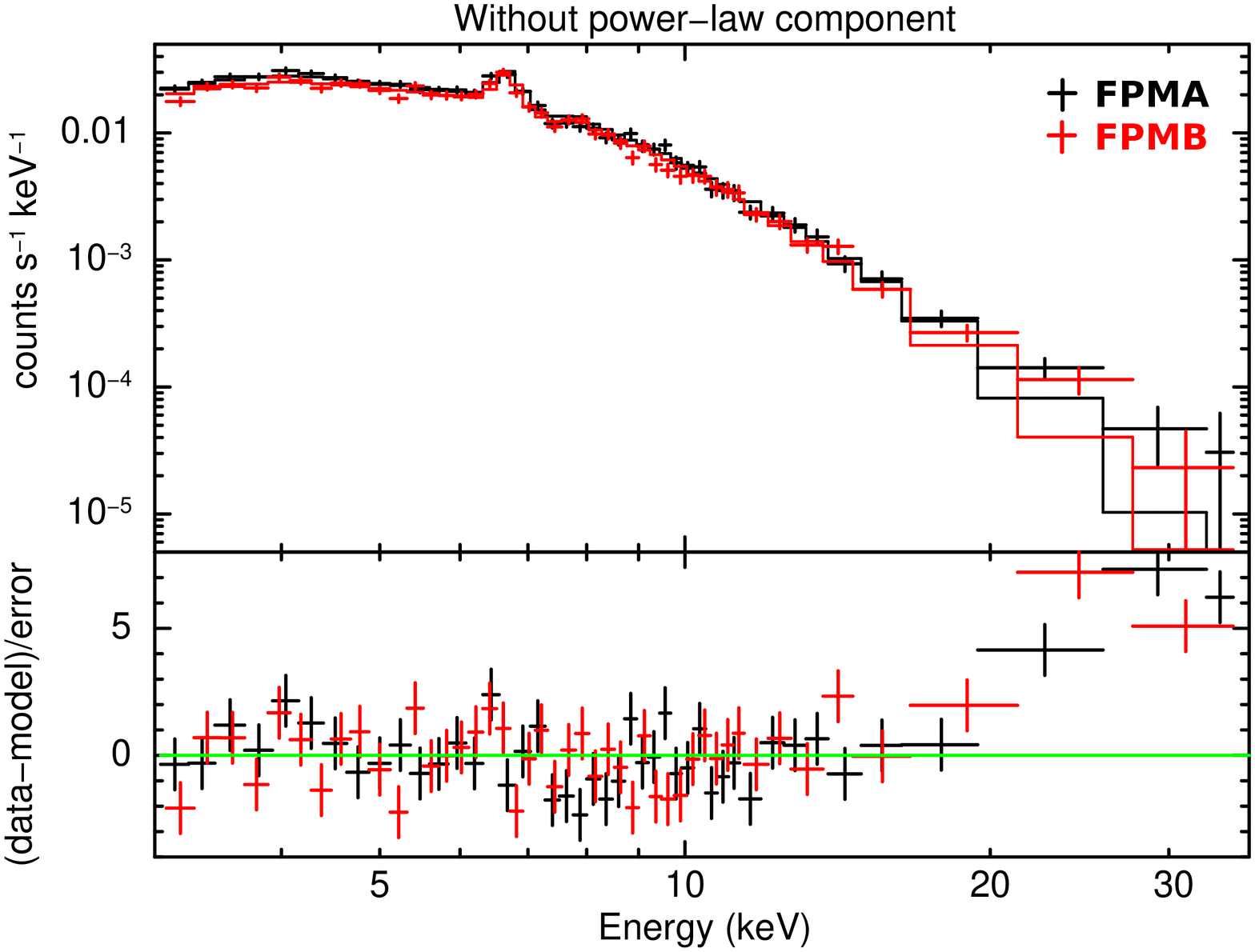}
    \includegraphics[width=0.49\linewidth]{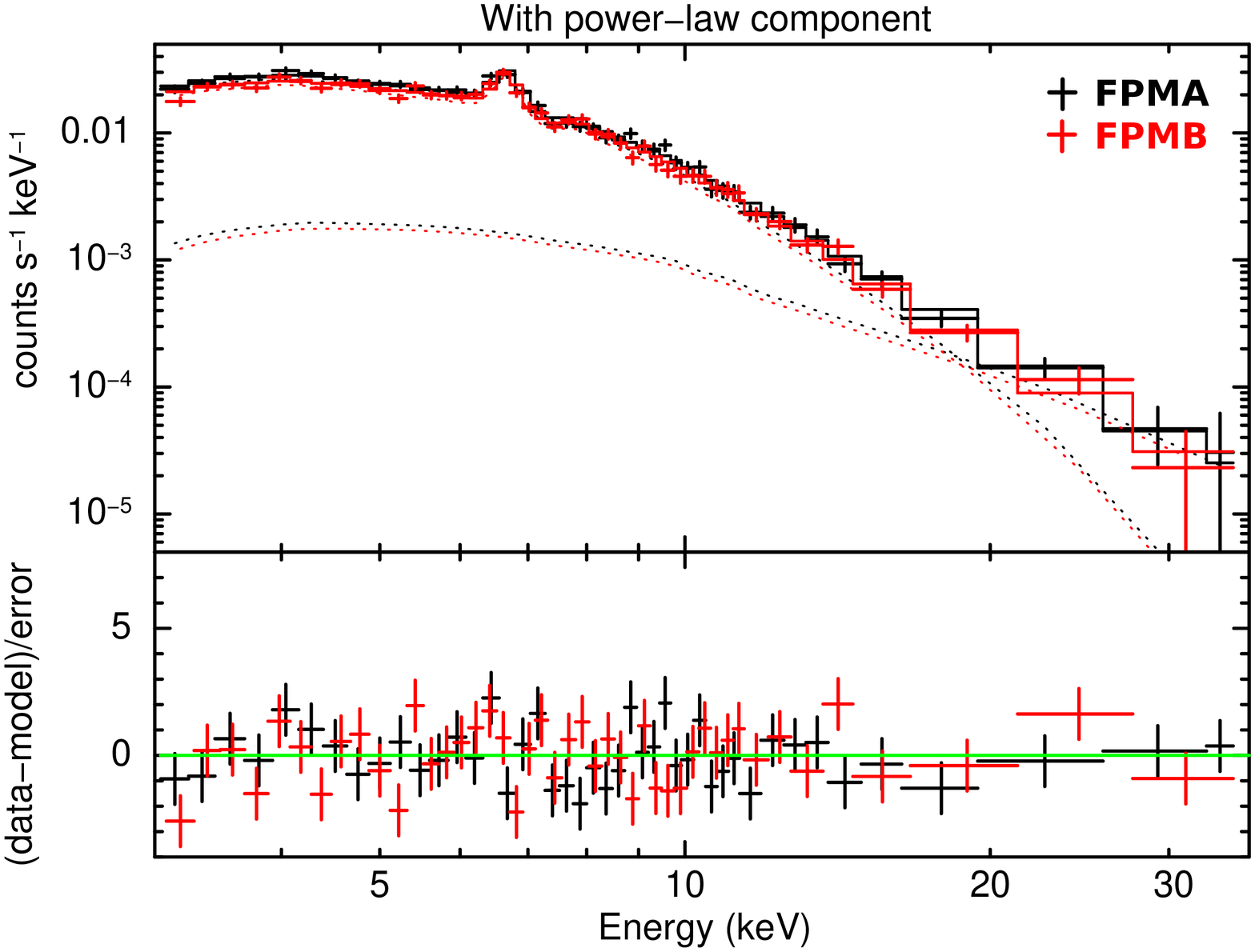}
    \caption{\textit{Apep} \Nu spectra in the 3--35~keV energy range from co-adding the observations for each camera separately. In the left panel the fitted model is \textit{TBabs*vphabs*vpshock}, while in the right panel an additional power-law component is added as \textit{TBabs*vphabs*(vpshock+po)}.}
    \label{fig:nustar_spectrum_fit}
\end{figure*}

We tried fitting different models to the \Nu spectra. The main conclusion is that these spectra are not sensitive to the adopted absorption model (as the absorption at energies $>3$\,keV is negligible) nor to the specifics of the emission model (as the information from emission lines is poor). Thus, we simply adopted the same model used to fit the \XMM spectra, \textit{constant*TBabs*vphabs*vpshock}, leaving the abundances and absorption fixed (but allowing temperature and normalisation to vary). Re-fitting the spectrum allowed us to obtain the normalisation constant between FPMA (taken as unity) and FPMB, $C=0.958\pm0.015$. A high-temperature ($k\,T \sim 5$~keV) thermal component naturally extends to energies above 10~keV and can explain most of the emission detected with \Nu. In Fig.~\ref{fig:nustar_spectrum_fit} we show the spectra and the residuals. However, the residuals at energies $> 20$\,keV become increasingly large, with deviations between 3--7\,$\sigma$. Such deviations can be attributed to the putative non-thermal component. We therefore included an additional \textit{power-law (po)} component. This way we obtained a significantly improved fit, as can be seen in the residuals in the right panel of Fig.~\ref{fig:nustar_spectrum_fit}, as well as in the lower C-stat value (which decreased from 237.5/182 to 203.1/181). For completeness, we note that an additional high-temperature \textit{vpshock} component with $kT > 10$~keV also leads to a similar improvement in the fit (although with higher residuals at energies $\sim 30$~keV). However, such high temperatures are not expected in CWB shocks, while a non-thermal component should arise naturally given the (already established in the radio band) presence of non-thermal particles. Thus, the main conclusion here is that the \Nu data by itself strongly supports the existence of an additional high-energy component, which we interpret hereon as a (non-thermal) power-law component. More robust results can be derived by including the data from \XMM consistently, which we address in the next section.

%-----------------------------------------------------
\subsection{Joint analysis}
%-----------------------------------------------------

The previous fitting to each instrument separately allowed us to understand the behaviour of the X-ray spectra and which observations are more sensitive to each component. Namely, the \XMM data allows us to better constrain the absorption and thermal emission models, whereas the \Nu data is sensitive to the putative high-energy non-thermal component. We now present a joint analysis of the whole data set for the same models as before. In general, we allowed for a different normalisation constant between the different instruments\footnote{Differences in absolute flux calibration between \XMM and \Nu can be of 5--15\% \citep{Madsen2017}.} and tied the remaining physical parameters. 

\begin{figure*}
    \centering
    \includegraphics[angle=0, width=0.492\linewidth]{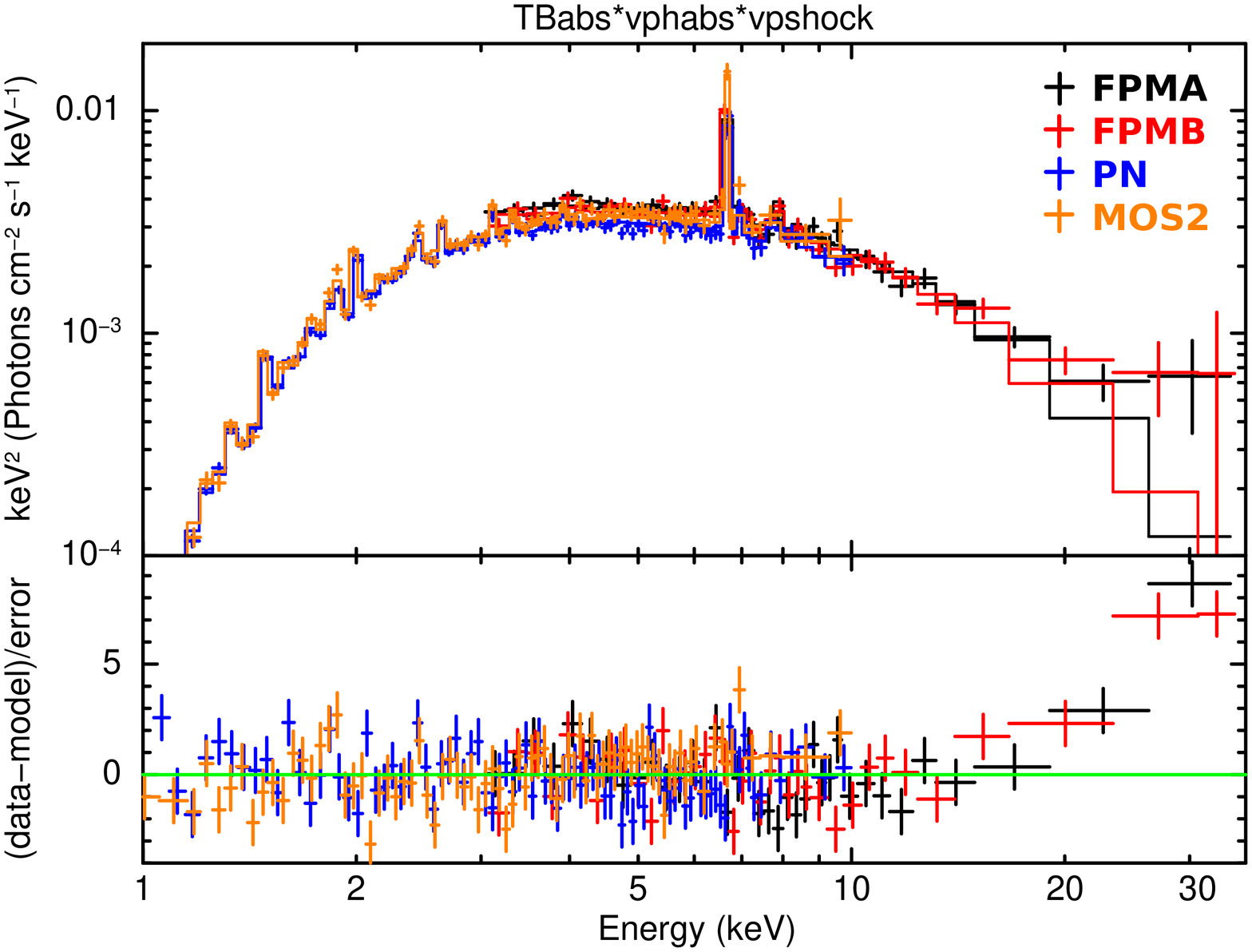} \hspace{0.1cm}
    \includegraphics[angle=0, width=0.492\linewidth]{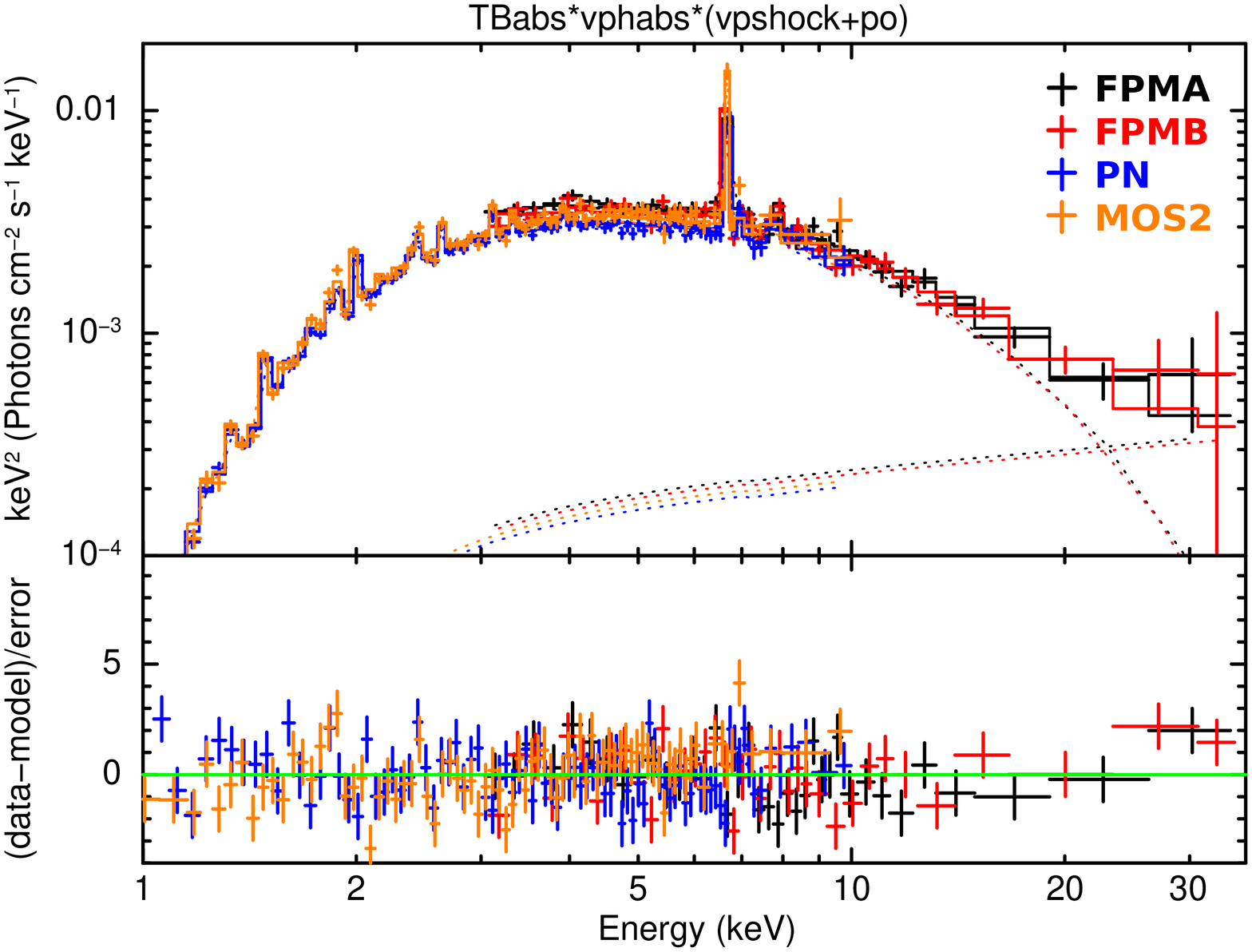}
    \caption{\textit{Apep} unfolded X-ray spectra in the 1--35~keV energy range with \XMM and \Nu. The fitted models are \textit{constant*TBabs*vphabs*vpshock} (left panel) and \textit{constant*TBabs*vphabs*(vpshock+po)} (right panel). The spectra has been rebinned for clarity. The fitted parameters are given in Table~\ref{tab:fit_parameters}.}
    \label{fig:nustar_xmm_spectrum}
\end{figure*}

In Fig.~\ref{fig:nustar_xmm_spectrum} we show the spectra and the combined fitting for two different models, while in Table~\ref{tab:fit_parameters} we detail the fitted parameters for the most relevant models. The \textit{TBabs*apec} model, used previously by \cite{Callingham2019}, fails to reproduce the spectra measured by both \XMM and \Nu. The \textit{TBabs*vphabs*vpshock} model satisfactorily fits the \XMM spectra, but it struggles with the \Nu spectra at energies above 20~keV (Fig.~\ref{fig:nustar_xmm_spectrum}). Finally, the \textit{TBabs*vphabs*(vpshock+po)} model can fit all the spectra simultaneously. In this case the C-stat diminished from 510.5/402 to 491.2/401, being the improvement more significant for \Nu at the expense of a slightly worse fit for \XMM (Table~\ref{tab:fit_parameters}). We further quantified the significance of the power-law component using the task \texttt{simftest} in \textit{XSPEC}. We ran 11\,000 simulations and obtained 
a probability $<0.01\%$ of the data being consistent with a model without the power-law component, which corresponds to a significance $>3.91\sigma$. We also note that the inclusion of the power-law component has little effect on the overall fit, mainly by diminishing slightly the temperature of the thermal component (from $\approx$5.3 to $\approx$4.9~keV; Table~\ref{tab:fit_parameters}). At last, we introduced a \textit{cflux} component to calculate both the total flux in the 10--30 keV band, $F_\mathrm{10-30\,keV} = (1.99\pm0.11)\times 10^{-12}$~\flux, and the one coming only from the power-law component, $F_\mathrm{10-30\,keV} = 4.8^{+1.0}_{-1.2}\times 10^{-13}$~\flux. 

We note that the flux of the power-law component is susceptible to the background extraction region chosen for \Nu, although its presence is always statistically favoured. A detailed exploration of different background regions and the caveats in their selection is given in Appendix~\ref{appendix:background}, together with a complementary analysis of the background using \texttt{nuskybgd} \cite{Wik2014}. The latter yields a flux of $F_\mathrm{10-30\,keV} \sim 3.9^{+1.0}_{-1.2}\times 10^{-13}$~\flux for the power-law component, which is consistent with the previous result within $1\sigma$. 

We also corroborated whether the value adopted for the power-law index had a significant impact on the results. For $\Gamma$ in the range 1.6--1.8, $F_\mathrm{10-30\,keV}$ varied only slightly ($\sim2\%$), much less than 1$\sigma$. Thus, the value adopted for $\Gamma$ does not affect the integrated flux in the hard X-ray band. We conclude that the presence of a power-law component is robust, and that its flux can only be measured with a rather high uncertainty.

\setlength{\tabcolsep}{4.0pt}
\begin{table*}
\caption{Results of the fitting of \Nu and \XMM spectra of \textit{Apep} using different models. C-statistics were used. The errors at 1-$\sigma$ level are specified for all parameters. A multiplicative constant was also added with a value set to 1 for FPMA and fitted for the other instruments to  $C_\mathrm{FPMB}=0.957\pm0.015$, $C_\mathrm{PN}=0.849\pm0.010$, and $C_\mathrm{MOS2}=0.902\pm0.011$. The best model is highlighted in boldface.}
\label{tab:fit_parameters}
\centering
\begin{tabular}{l c c c c}
\hline \hline\\[-7pt] 
 Parameter   & Units                & \textit{TBabs*apec}       &  \textit{TBabs*vphabs*vpshock}  &   \textbf{\textit{TBabs*vphabs*(vpshock+po)}}  \\
\hline\\[-7pt] 
$N_{\rm H}$ & $10^{22}$~cm$^{-2}$   & $2.67\pm0.02$             &  1.63 (fixed), $1.23^{+0.06}_{-0.06}$    &  1.63 (fixed), $1.20^{+0.07}_{-0.05}$   \\[2pt]
$k\,T$      & keV                   & $5.15\pm0.07$             &  $5.27^{+0.07}_{-0.06}$   &  $4.91\pm0.11$   \\[2pt] 
norm$_1$    & $10^{-2}$~cm$^{-5}$   & $1.102^{+0.020}_{-0.019}$ &  $1.023^{+0.018}_{-0.017}$&  $1.014^{+0.017}_{-0.019}$\\[2pt] 
$A$         &                       & $0.52\pm0.02$             &  ---                      &  ---                      \\[2pt] 
Fe          &                       &   ---                     &  $0.54\pm0.02$            &  $0.58\pm0.02$\\[2pt] 
Ne          &                       &   ---                     &  $2.88^{+0.38}_{-0.34}$   &  $3.36^{+0.50}_{-0.44}$  \\[2pt] 
S           &                       &   ---                     &  $0.47\pm0.07$            &  $0.43^{+0.08}_{-0.06}$  \\[2pt] 
Ca          &                       &   ---                     &  $0.58^{+0.23}_{-0.21}$   &  $0.57^{+0.20}_{-0.22}$ \\[2pt] 
$\tau_\mathrm{u}$ & $10^{11}$\,s\,cm$^{3}$ & ---                &  $9.01^{+0.96}_{-0.86}$   &  $9.36^{+1.31}_{-0.90}$  \\[2pt] 
$\Gamma$    &                       & ---                       &  ---                      &  1.72 (fixed) \\[2pt] 
norm$_2$    & 10$^{-4}$~keV$^{-1}$\,cm$^{-2}$\,s$^{-1}$  & ---  &  ---                      &  $1.30^{+0.26}_{-0.33}$   \\[2pt] 
\hline\\[-7pt] 
$C_\mathrm{stat}$/bins (PN)  &   & 234.3/112                  & 136.6/112                 & 137.4/112    \\[2pt]
$C_\mathrm{stat}$/bins (MOS2)&   & 206.9/115                  & 134.6/115                 & 141.0/115    \\[2pt]
$C_\mathrm{stat}$/bins (FPMA)&   & 125.7/91                   & 124.8/91                  & 112.1/91    \\[2pt]
$C_\mathrm{stat}$/bins (FPMB)&   & 122.6/95                   & 114.5/95                  & 101.4/95    \\[2pt]
\hline\\[-7pt] 
\textbf{$\mathbf{C_\mathrm{stat}}$/d.o.f. (total)}&  & 689.5/406  & 510.5/402 & \textbf{491.2/401}    \\[2pt]
\hline
\end{tabular}
\end{table*}

%===================================================================
\section{Discussion}
%===================================================================

The main result from our spectral analysis is the detection of hard X-ray emission consistent with a power-law component with a flux of $F_\mathrm{10-30\,keV} = 4.8^{+1.0}_{-1.2}\times 10^{-13}$~\flux. 
We also constrain the thermal emission from the WCR much better. Previous estimates of the plasma temperature by \cite{Callingham2019} were highly uncertain, spanning a range $kT \sim 4.7$--$6.3$~keV (for the particular \XMM observation that we also analysed, they obtained $kT \sim 4.9$--5.3~keV). By means of a more careful analysis of the \XMM observations, combined with the unique information provided by \Nu above 10 keV, we constrain it to $kT \approx 4.85$--$5.07$~keV.

We now focus on the interpretation of the non-thermal component. To be more conservative, in what follows we consider additional sources of errors in the flux values, both statistical and systematic. For example, a 10\% systematic error due to absolute calibration uncertainties are estimated for \Nu \citep{NuSTAR2013}. There is also an additional dependency of the retrieved flux with the chosen background region, as shown in Appendix~\ref{appendix:background}. Based on this, we adopt a less constrained flux of $F_\mathrm{10-30\,keV} = (2$--$6)\times 10^{-13}$~\flux in our analysis, which is slightly broader than the 90\% confidence interval obtained from the analysis with \texttt{nuskybgd} in Appendix~\ref{appendix:background} ($F_\mathrm{10-30\,keV} \sim  (2.3$--$5.6)\times 10^{-13}$~\flux).

%------------------------------------------------------------------
\subsection{Modelling the hard X-ray emission}\label{sec:X-R_model}

According to \cite{delPalacio2022}, the hard X-ray emission should arise from IC scattering of stellar photons by relativistic electrons in the WCR. These same electrons should also be responsible for the non-thermal (synchrotron) emission observed in the radio band \citep{Callingham2019, Marcote2021, Bloot2022}. In order to relate the observed fluxes with the particle acceleration in the shocks, we need to take into account that only a fraction of the wind kinetic power can be converted into relativistic particles in the shocks, that this energy is distributed into electrons and protons, and that each of this particle species radiate only a fraction of their energy at any given frequency range \citep[e.g.][]{DeBecker2013}. 

Thus, to model this emission, we use a code based on the non-thermal emission model presented in \cite{delPalacio2016} with the system parameters listed in Table~\ref{tab:system_parameters}. The model solves for the acceleration and transport of relativistic particles for both shocks in the WCR (one for each stellar wind). These relativistic particles radiate by different processes (synchrotron, IC, p-p collisions), and this radiation is mitigated by absorption processes in the stellar winds or radiation fields. %, all of which is computed consistently. 
The model has two free parameters that determine the leptonic emission: the ratio between the magnetic field pressure to thermal pressure in the WCR, $\eta_B$, and the fraction of the available power at the shocks that is converted into relativistic electrons, $f_\mathrm{NT,e}$. The available power for particle acceleration is the wind kinetic power injected perpendicularly into the WCR shocks. Denoting this power by $L_\mathrm{inj,\perp}$ and the total wind kinetic power of a star by $L_\mathrm{w}=0.5 \dot{M} v_\mathrm{w}^2$, we can write $L_\mathrm{inj,\perp} = \epsilon L_\mathrm{w}$, with $\epsilon_\mathrm{WN}=8\%$ and $\epsilon_\mathrm{WC}=18\%$ for the system \textit{Apep}. Here, the value of $\epsilon$ depends on the geometry of the WCR, which is governed by the value of the wind-momentum rate ratio $\eta$, and is calculated numerically in the model. Further details of the model are described in Appendix~\ref{sec:NT_model}. 

\begin{figure}
    \centering
    \includegraphics[width=\linewidth]{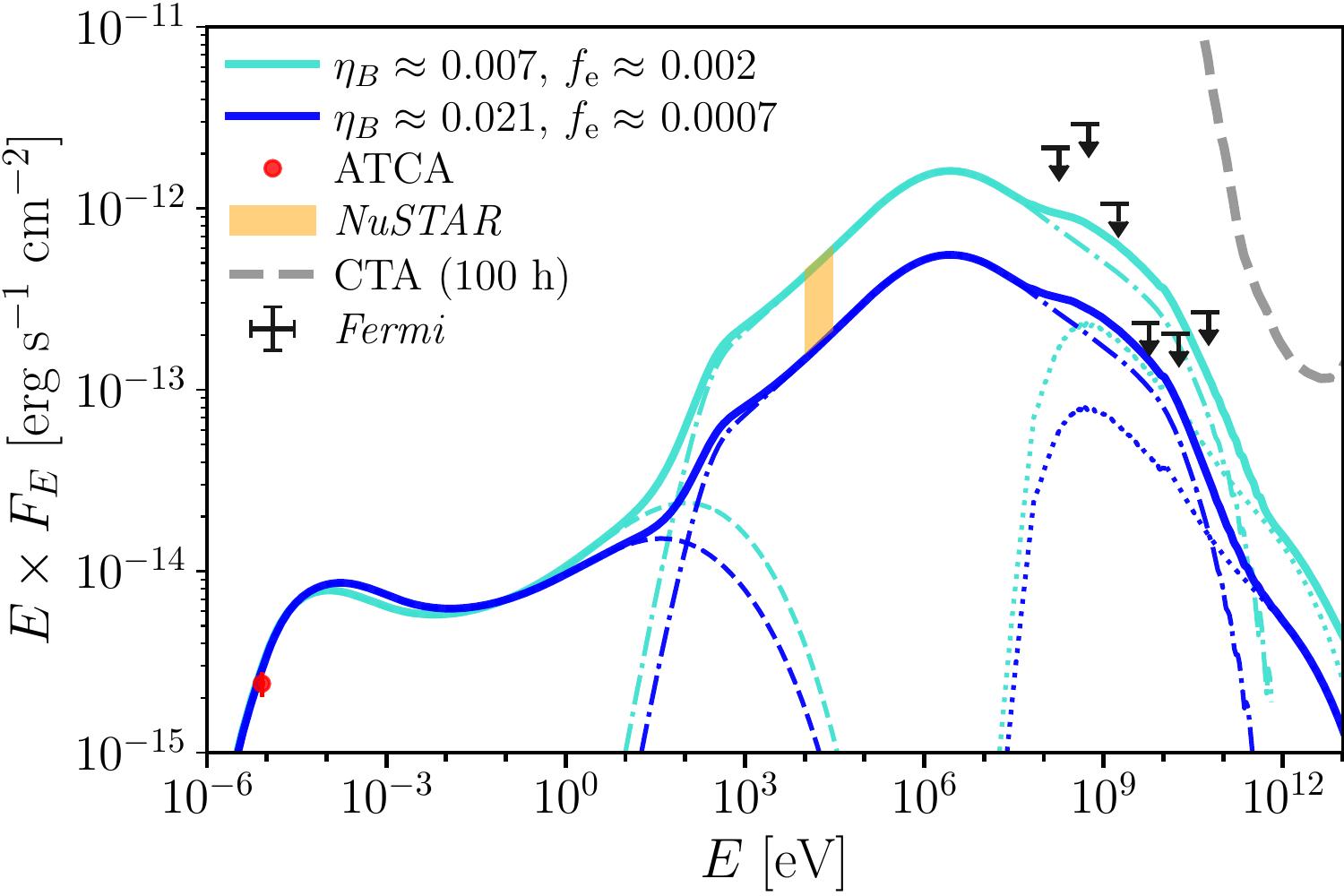}
    \caption{Modelled broadband non-thermal SED of \textit{Apep}. We show SEDs for the two extreme combinations of parameters that are compatible with the hard X-ray emission component detected with \Nu (this work) and with the ATCA radio data \citep{Callingham2019}. Dashed lines show the synchrotron component, dot-dashed lines the IC component, dotted lines the p-p component, and solid lines the total SED. We also show the \textit{Fermi}-LAT upper-limits from \cite{Marti-Devesa2022} 
    and the sensitivity curve for CTA \citep{Funk2013}.}
    \label{fig:SED_model}
\end{figure}

It is possible to tie the two free parameters, $f_\mathrm{NT,e}$ and $\eta_B$, by modelling the observed synchrotron component \citep{delPalacio2016}. In this case the relation is $f_\mathrm{NT,e} \eta_B = constant$ \citep{delPalacio2020}. However, it is not possible to break the degeneracy between these two parameters from radio data alone. Fortunately, this can be solved when observations in hard X-rays measure the flux from the IC component, which depends only on $f_\mathrm{NT,e}$ as $F_\mathrm{IC} \propto f_\mathrm{NT,e}$ \citep{delPalacio2020}. 
Once $f_\mathrm{NT,e}$ is derived from the measured hard X-ray flux, we can obtain $\eta_B$ by fitting the synchrotron emission to the observed radio flux of $\approx 120$~mJy at 2~GHz \citep{Callingham2019}. In Fig.~\ref{fig:SED_model} we show the SEDs fitted using this procedure. In addition, for a given value of $\eta_B$ we can calculate the magnetic field in the apex of the WCR, $B_\mathrm{WCR}$. 

Previous estimates by \cite{delPalacio2022} based only on the synchrotron emission from the source had more than one order of magnitude of uncertainty in $f_\mathrm{NT,e}$, namely $f_\mathrm{NT,e} \approx (0.11$--$2.7)\times 10^{-3}$. 
Our new estimates based on the \Nu detection yield a very well-constrained value with less than a factor two uncertainty, $f_\mathrm{NT,e} \approx (0.7$--$2)\times 10^{-3}$.
This corresponds to roughly $1.5\times10^{-4}$ of the total wind kinetic power being converted into relativistic electron acceleration. 
Moreover, the magnetic field in the WCR was also poorly constrained by \cite{delPalacio2022} to $B_\mathrm{WCR} \approx 70$--400~mG ($\eta_B = (3$--$100)\times 10^{-3}$), while now we constrained it to $B_\mathrm{WCR} \approx 105$--190~mG ($\eta_B = 0.007$--$0.021$). This translates into a ratio between the energy density in relativistic electrons and the magnetic field of $U_\mathrm{e}/U_B \approx 0.02$--0.2. For reasonable values of a ratio between power injected in electrons and protons of $K_\mathrm{e,p} < 0.1$, this leads to a magnetic field in subequipartition with the non-thermal particles, supporting the possibility of relativistic protons driving the magnetic field amplification \citep{Bell2004}.

These values can be compared with those found for the O+O binary HD\,93129A during its periastron passage \citep{delPalacio2020}, $f_\mathrm{NT,e} \approx 6\times 10^{-3}$ and $\eta_B \sim 0.02$ ($B_\mathrm{WCR} \approx 0.5$~G). Comparisons with other systems are complicated given the uniqueness of the detection of a PACWB in both radio and high energies. Nonetheless, we can comment on the PACWB $\eta$-Car, for which non-thermal hard X-rays were also detected with a power-law index $\Gamma \sim 1.65$ \citep[although poorly constrained;][]{Hamaguchi2018}. Another systems studied by \cite{DeBecker2018}, based on radio observations and equipartition assumptions between relativistic particles and magnetic fields, yielded that the fraction of the wind kinetic power converted into relativistic electrons is $\sim 10^{-4}$--$10^{-6}$ for Cyg~OB2~\#8a, $\sim 10^{-7}$--$10^{-9}$ for WR~140, and $\sim 10^{-5}$--$10^{-7}$ for HD~167971. Compared with the value obtained here for \textit{Apep} ($1.5\times10^{-4}$), it is clear that this system is a much more efficient electron accelerator. This is consistent with the fact that this binary is the brightest synchrotron-emitting PACWB. 
We also tried to compare the values of $\eta_B$ with those derived from \cite{DeBecker2018}. These values are $10^{-7}$ for WR~140, $2\times10^{-4}$ for Cyg~OB2~\#8a, and $5\times10^{-5}$ for HD~167971, but these have an uncertainty that spans 2--3 orders of magnitude, so all we can say is that our value of $\eta_B \approx 10^{-2}$ is exceptionally well-constrained. In addition, \cite{Pittard2021} fitted the radio SED of the system WR~146 to derive a magnetic field compatible with $\eta_B \approx 10^{-3}$, although these authors required a very large particle efficiency in return ($f_\mathrm{NT} \approx 0.3$). 

One last parameter we could derive from our results is the surface magnetic field of the stars. For this we assumed a toroidal stellar magnetic field that drops as $r^{-1}$ and is adiabatically compressed in the WCR shocks, and a stellar rotation velocity of $V_\mathrm{rot} \sim 0.1 v_\infty$ \citep[][and references therein]{delPalacio2016}. Under these assumptions we obtain values of the surface stellar magnetic fields in the ranges $B_\mathrm{WN} = 650$--1100~G and $B_\mathrm{WC} = 280$--490~G. Nonetheless, it is possible that magnetic field amplification processes take place in the WCR shocks \citep[e.g.][]{Bell2004,Pittard2021}. In this case the aforementioned values should actually be interpreted as upper limits to the stellar magnetic fields.

%------------------------------------------------------------------
\subsection{Predictions of $\gamma$-ray emission}\label{sec:gamma-rays}

The previous estimates of the power in relativistic electrons also allowed us to compute the expected IC luminosity in the $\gamma$-ray domain. In Fig.~\ref{fig:SED_model} we show the modelled broadband SED extending to $\gamma$-ray energies, together with characteristic sensitivity thresholds of $\gamma$-ray observatories. We first focus on the 0.1--100~GeV energy range, which can be tested with observations with the \textit{Fermi}-LAT instrument. The $\gamma$-ray luminosity in this case is $F_\mathrm{0.1-100\,GeV} = (1.9\pm0.9)\times 10^{-12}$\,\flux, though it can be larger if a hadronic component is included (for example, $K_\mathrm{e,p}=0.04$ yields to a total flux of \mbox{$F_\mathrm{0.1-100\,GeV} = (2.0\pm1.5)\times 10^{-12}$\,\flux}). These values are mostly consistent with a non-detection of this source with \textit{Fermi}-LAT at a level of $F_\mathrm{0.1-100\,GeV} \sim (1$--$2)\times 10^{-12}$\,\flux \citep{Marti-Devesa2022}. The small tension between the higher fluxes predicted for the lower magnetic field scenarios might suggest that the higher magnetic field scenarios are to be preferred (Fig.~\ref{fig:nustar_spectrum_fit}). Nonetheless, this tension can also be attributed to even small uncertainties in the particle energy distribution that lead to a significant difference in the predicted $\gamma$-ray fluxes. Assuming that the injected electron energy distribution is slightly harder, $p=2.3$ (equivalently, \mbox{$\Gamma=1.65$}), we obtain \mbox{$F_\mathrm{0.1-100\,GeV} \sim 4.2\times 10^{-12}$\,\flux}, while a slightly steeper distribution, \mbox{$p=2.55$} ($\Gamma=1.77$), yields an IC emission of \mbox{$F_\mathrm{0.1-100\,GeV} \sim (1.0\pm0.7)\times 10^{-12}$\,\flux}. Thus, a hardening of the electron energy distribution is strongly disfavoured as it would overpredict the $\gamma$-ray luminosity. Moreover, we conclude that the source is either close to be detected by \textit{Fermi}, or that a non-detection with deeper sensitivity (\mbox{$F_\mathrm{0.1-100\,GeV} < 10^{-12}$\,\flux}) would mean that the SED softens at energies above hard X-rays, which in turn would require a softening in the electron energy distribution at energies $E_\mathrm{e} > 100$~MeV. % The photon energy is E_gamma ~ E_star * gamma_e^2, with E_star ~ 10 eV, so for E_gamma ~ 100 keV we get gamma_e ~ 100. 

Finally, we address the prospects for detection of TeV emission from \textit{Apep}. We predict an IC flux of \mbox{$F_\mathrm{TeV} \sim 1.8\times 10^{-14}$\,\flux} in the 0.1--10~TeV energy range, although the poorly-constrained hadronic component is likely dominant \citep{delPalacio2022}: assuming $K_\mathrm{e,p}=0.04$, the predicted total flux (p-p + IC) is $F_\mathrm{TeV} \sim 8\times 10^{-14}$\,\flux. Moreover, small variations in the spectral index of the particle energy distribution ($p=2.3$--2.5) can lead to significantly different TeV fluxes, $F_\mathrm{TeV} \sim (0.3$--$18)\times 10^{-14}$\,\flux. Only the higher fluxes would potentially be detectable by the Cherenkov Telescope Array \citep[CTA;][]{Funk2013}, but these are already disfavoured in view of the lack of detections of GeV emission. Thus, the TeV emission from \textit{Apep} seems too faint to be detected with current and upcoming TeV observatories.

%===================================================================
\section{Conclusions}
%===================================================================

We presented the first hard X-ray view of the PACWB \textit{Apep}. This system is the brightest synchrotron source among the known PACWBs. The \Nu spectrum revealed strong evidence of a power-law component, consistent with the predicted IC emission produced by relativistic electrons in the WCR. The detection of this non-thermal high-energy emission from a system that also presents non-thermal emission in the radio band represents an observational breakthrough in the study of PACWBs. In particular, it has allowed us to place the tightest constraints to the magnetic field and electron acceleration efficiency in the WCR of a PACWB. We also predict that \textit{Apep} is close to be detected at $\gamma$-rays with \textit{Fermi} unless the electron energy distribution softens at energies $>100$~MeV. 

We highlight the importance of multi-wavelength observations for improving our understanding of PACWBs. Unfortunately, the high-energy emission from these systems is rather weak and difficult to detect, but at least for the brightest sources observations in the hard X-ray and high energy $\gamma$-rays bands have proven to be successful, paving the way for moving forward in the research of PACWBs.

%===================================================================

\begin{acknowledgements}
We would like to thank the referee for providing useful feedback that helped us to improve the manuscript. 
This work was carried out in the framework of the PANTERA-Stars\footnote{\url{https://www.astro.uliege.be/~debecker/pantera/}} initiative. FG is CONICET Researcher and was supported by PIP 0113 (CONICET), PICT-2017-2865 (ANPCyT), and PIBAA 1275 (CONICET). FG was also supported by grant PID2019-105510GB-C32/AEI/10.13039/501100011033 from the Agencia Estatal de Investigaci\'on of the Spanish Ministerio de Ciencia, Innovaci\'on y Universidades, and by Consejer\'{\i}a de Econom\'{\i}a, Innovaci\'on, Ciencia y Empleo of Junta de Andaluc\'{\i}a as research group FQM-322, as well as FEDER funds.
D.A. acknowledges support from the Royal Society.
V.B-R. is Correspondent Researcher of CONICET, Argentina, at the IAR. GER was supported by grant PIP 11220200100554CO (CONICET).
This work received financial support from the State Agency for Research of the Spanish Ministry of Science and Innovation under grant PID2019-105510GB-C31/AEI/10.13039/501100011033 and through the ''Unit of Excellence Mar\'ia de Maeztu 2020-2023'' award to the Institute of Cosmos Sciences (CEX2019-000918-M). 
This work also made use of the softwares \texttt{ds9} \citep{ds9_2003} and \texttt{Matplotlib} \citep{Matplot_2007}.

\end{acknowledgements}

\bibliographystyle{aa} 
\bibliography{references} 
%-------------------------------------------------------------------

%%%%%%%%%%%%%%%%%%%%%%%%%%%%%%%%%%%%%%%%%%%%%%%%%%

%%%%%%%%%%%%%%%%% APPENDICES %%%%%%%%%%%%%%%%%%%%%

\appendix

\section{\Nu co-added data} \label{appendix:co-add}

\begin{figure}
    \centering
    \includegraphics[width=\linewidth]{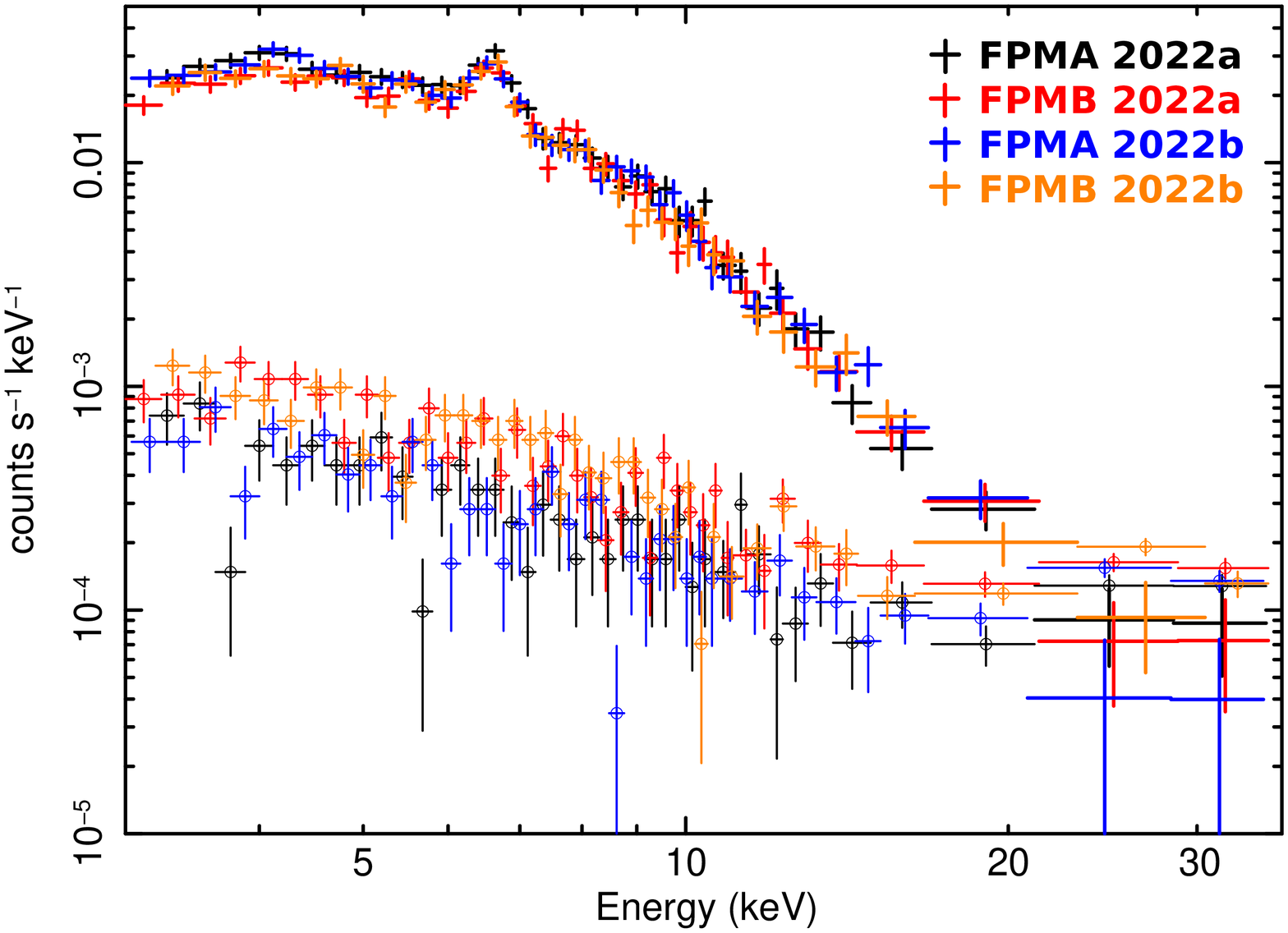}\\
    \includegraphics[width=\linewidth]{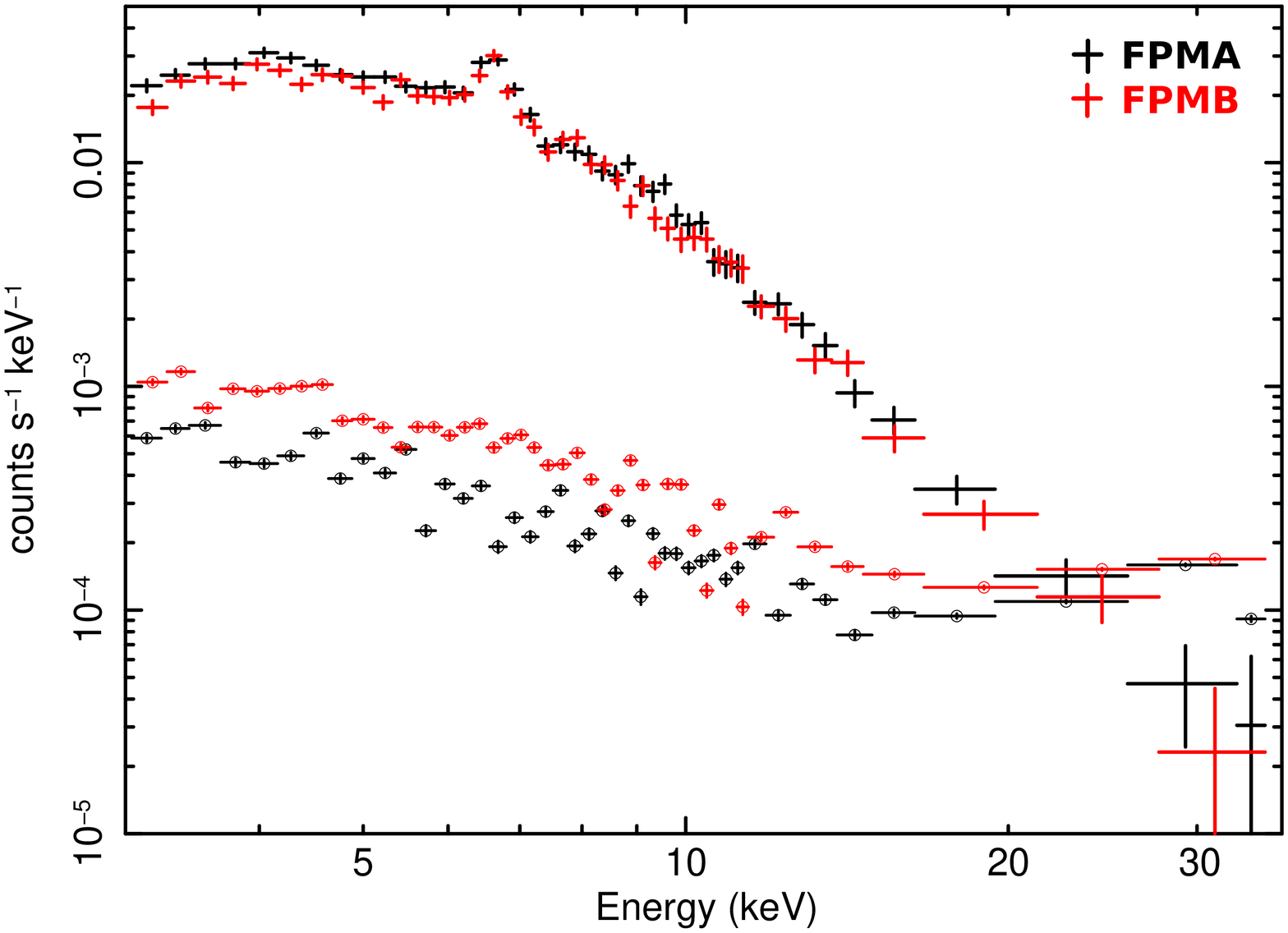}\\
    \includegraphics[width=\linewidth]{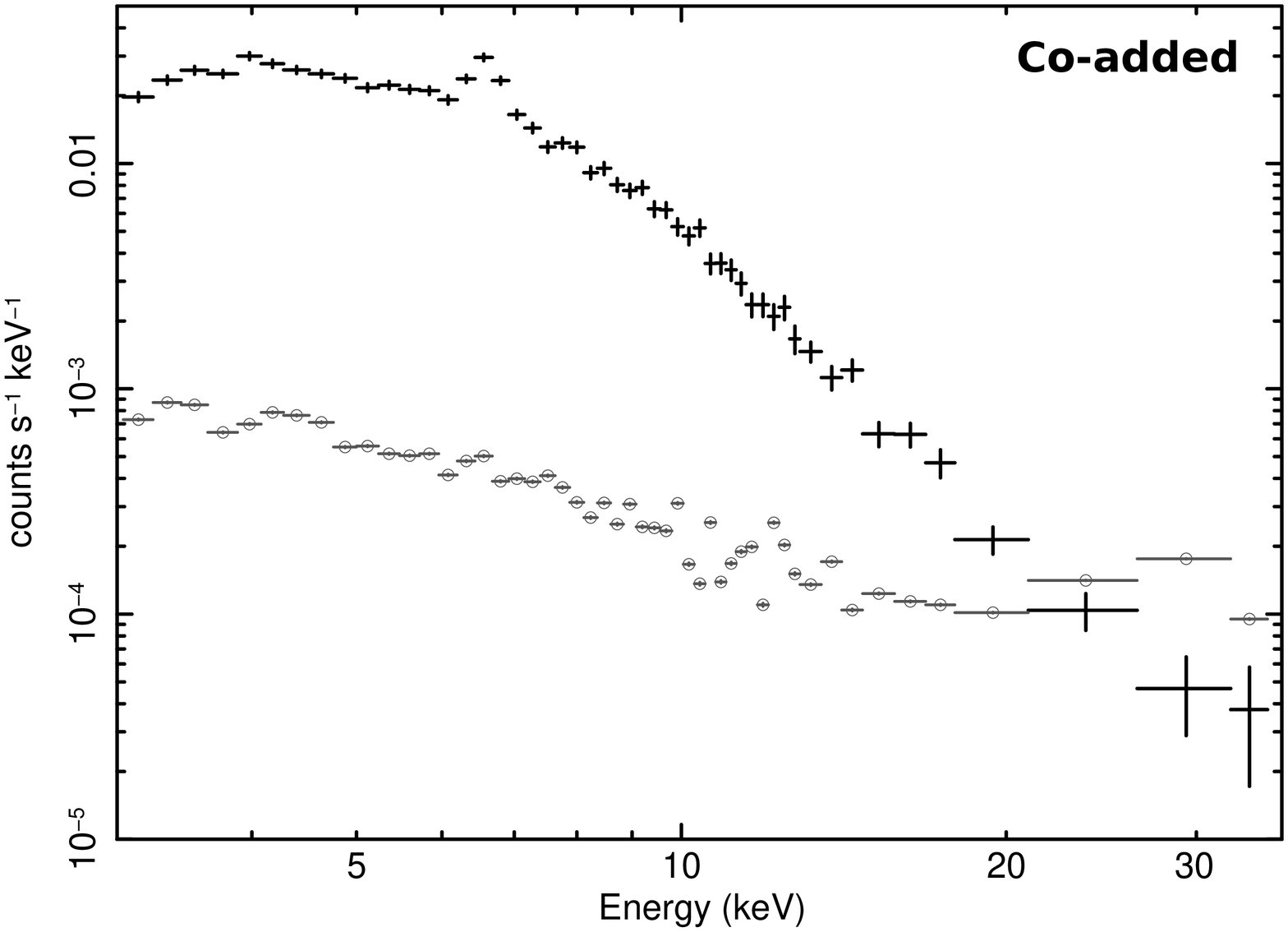}
    \caption{\textit{Apep} \Nu spectra in the 3--35~keV energy range. The spectra was further rebinned in \texttt{XSPEC} for clarity. The source spectra is shown with crosses and the background with open circles. \textit{Top panel:} spectra for the 2022a and 2022b observations showing each camera separately. \textit{Middle panel:} spectrum from the co-added observations from both epochs (for each camera separately).  \textit{Bottom panel:} spectrum from the co-added observations (both epochs, both cameras).}
    \label{fig:nustar_spectrum_compare}
\end{figure}

In the top panel of Fig.~\ref{fig:nustar_spectrum_compare} we show the \Nu spectra for both \Nu observations and both cameras independently. As detailed in Sect.~\ref{sec:nustar}, we co-added the spectra of both epochs for each camera to compare if there was any systematic difference between FPMA and FPMB. The result is shown in the middle panel of Fig.~\ref{fig:nustar_spectrum_compare}\footnote{We note that the apparently very small background errorbars returned by the \texttt{addspec} task are appropriate (FTOOLS helpdesk, priv. comm.).}. Even though both cameras are compatible, there is a small difference between them at energies $<10$~keV, and for that reason they were not combined in the detailed analysis carried out in Sect.~\ref{sec:results}. Nonetheless, we also explored the result of co-adding both cameras in order to improve the signal-to-noise at the highest energies (neglecting the error introduced at lower energies). The result is shown in the bottom panel of Fig.~\ref{fig:nustar_spectrum_compare}. This last spectrum has improved statistics at energies $>20$~keV and allows us to better see by eye the shape of the SED. Additionally, the fitting of this co-added dataset retrieved fully compatible results on the best-fitting parameters to the ones obtained in Sect.~\ref{sec:nustar} without co-adding the cameras. 

%-------------------------------------------------
\section{\Nu background} \label{appendix:background}

As discussed in Sect.~\ref{sec:results}, the selection of the background extraction region can have a mild impact in the high-energy spectrum of \textit{Apep} derived with \Nu. We explore this in more detail to make sure that our results are robust. 

\begin{figure}
    \centering
    \includegraphics[width=\linewidth]{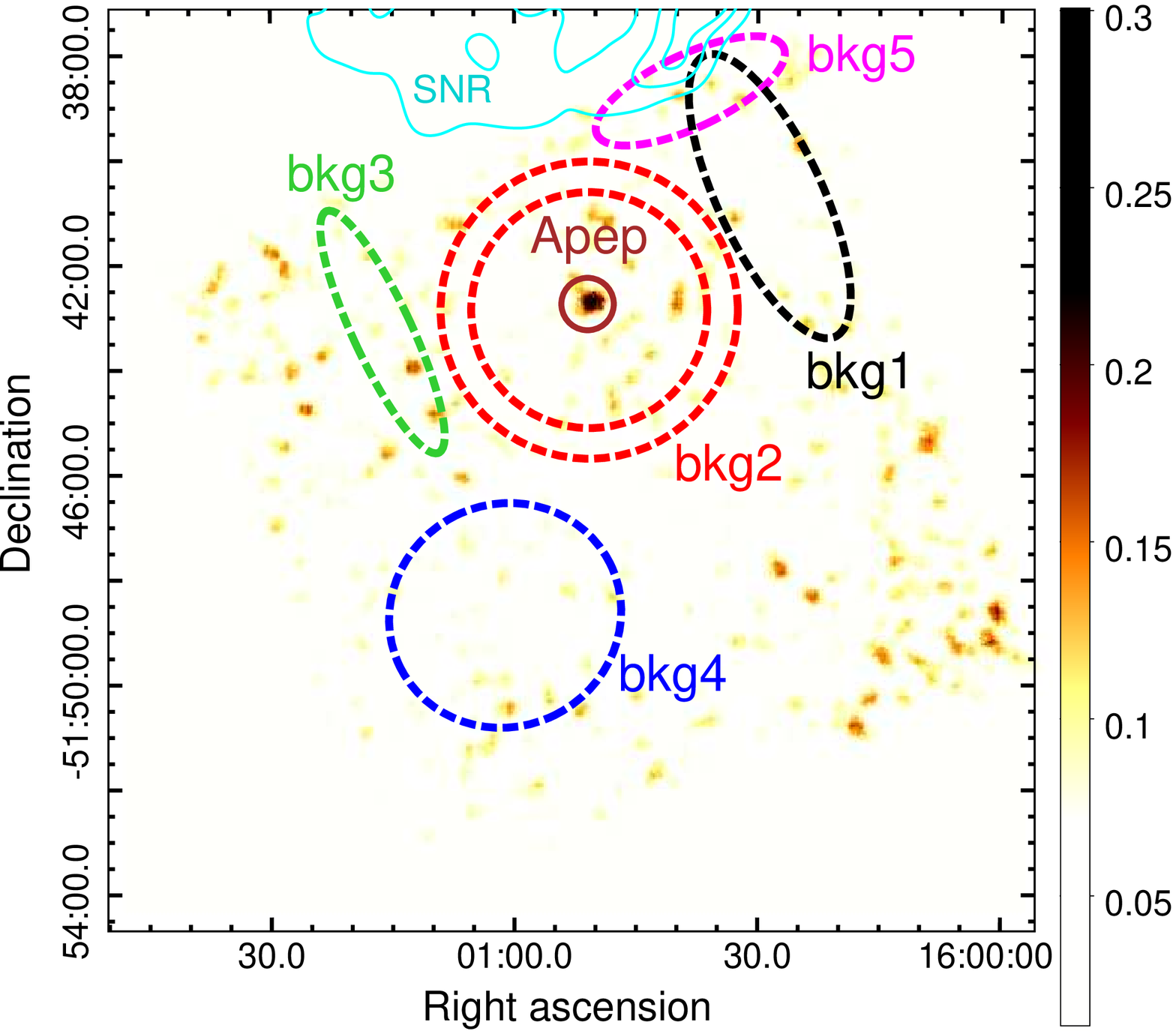}\\
    \includegraphics[width=\linewidth]{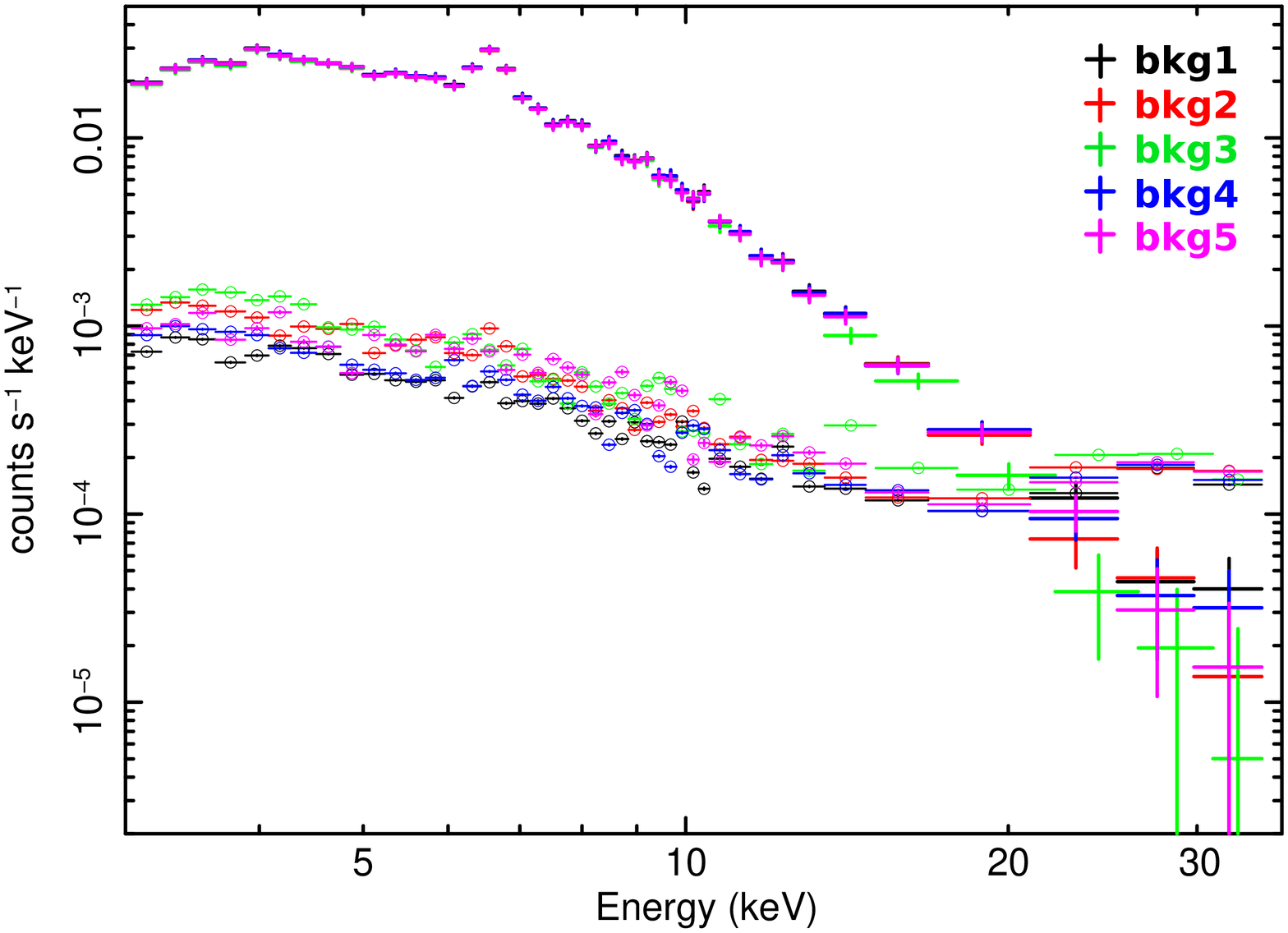}
    \caption{\textit{Top:} \Nu image in the 20--35~keV energy range for FPMB (2022a). We mark the position of \textit{Apep} with a 30\arcsec-circle, the background extraction region used in the analysis (bkg1), and the alternative backgrounds used as independent checks (bkg2--bkg5). We also show in cyan contours the emission from the supernova remnant G330.2+1.0 obtained in the 2.5--8~keV range with \XMM. The image is in linear scale (instead of logarithmic scale as in Fig.~\ref{fig:nustar_image}) and it was smoothed in \texttt{ds9} using a Gaussian with \texttt{Radius=5}. \textit{Bottom:} 
    Background-subtracted spectra (crosses) and background spectra (open circles) of \textit{Apep} for the different background regions considered in this Appendix B.
    In all cases the whole dataset was co-added and rebinned to increase the S/N and ease the visual comparison.}
    \label{fig:backgrounds}
\end{figure}

In Fig.~\ref{fig:backgrounds} (top panel) we show an image in the 20--35~keV range of the \textit{Apep} field of view for the first observation with FPMB, in which \textit{Apep} is clearly detected above the local background. We note that other high-intensity background regions appear in this image, but these are not consistent in both observations and FPM cameras; on the contrary, the count excess at the position of \textit{Apep} is confirmed in all cases, strongly supporting its detection. In this image we also mark the different background regions analysed here, labelled as \texttt{bkg1}--\texttt{bkg5}. The corresponding background-subtracted spectra of \textit{Apep} and the background spectra are also shown in Fig.~\ref{fig:backgrounds} (bottom panel).  

\begin{itemize}
    \item \texttt{bkg1}: This elliptical region is our preferred choice as it is within the same chip as the source, which is the usual recommendation for analysing \Nu data, and it is sufficiently far from \textit{Apep} and other sources in the field. This background region leads to a power-law component with a significance of $>3.91\sigma$ ($<0.01\%$ of by-chance detection) with a flux of $F_\mathrm{10-30\,keV} = 4.8^{+1.0}_{-1.2}\times 10^{-13}$~\flux.

    \item \texttt{bkg2}: An annulus region centred at the position of \textit{Apep} and with inner radius of 135\arcsec and outer radius of 170\arcsec. We note that $\approx90\%$ of the photons from \textit{Apep} should be contained within a 130\arcsec from the source; nonetheless, the presence of weak Fe line emission in the bkg spectrum (seen in Fig.~\ref{fig:backgrounds}, bottom panel) suggests a non-negligible contamination from the source (at least at energies below 10 keV). This background region leads to a power-law component with a significance of $2.12\sigma$ ($3.4\%$ of by-chance detection) with a flux of $F_\mathrm{10-30\,keV} = 2.3^{+1.1}_{-1.2}\times 10^{-13}$~\flux.  
    
    \item \texttt{bkg3}: This is an elliptical region within the chip with the highest background level (also appreciable in Fig.~\ref{fig:nustar_image}). This background region leads to a power-law component with a significance of $1.66\sigma$ ($9.7\%$ of by-chance detection) with a flux of $F_\mathrm{10-30\,keV} = 1.3^{+1.1}_{-1.0}\times 10^{-13}$~\flux.  
    
    \item \texttt{bkg4}: This is a circular region in another . This background region leads to a power-law component with a significance of $3.4\sigma$ ($0.07\%$ of by-chance detection) with a flux of $F_\mathrm{10-30\,keV} = 3.5^{+1.0}_{-1.1}\times 10^{-13}$~\flux.
    
    \item \texttt{bkg5}: This elliptical region is also within the same chip as the source, but we note that it is likely contaminated by the hard-spectra supernova remnant G330.2+1.0 \citep{Park2009} shown in Fig.~\ref{fig:backgrounds}. This background region leads to a power-law component with a significance of $>3.85\sigma$ ($0.012\%$ of by-chance detection) with a flux of $F_\mathrm{10-30\,keV} = 4.3^{+1.0}_{-1.2}\times 10^{-13}$~\flux.

    \item At last, we considered all of the previous regions altogether in order to have an estimate of an average background. This leads to a power-law component with a significance of $2.53\sigma$ ($1.1\%$ of by-chance detection) with a flux of $F_\mathrm{10-30\,keV} = 2.8^{+1.1}_{-1.2}\times 10^{-13}$~\flux.
    
\end{itemize}

We emphasise that the alternative backgrounds \texttt{bkg2}--\texttt{bkg4} analysed here do not lie (at least completely) within the same chip as the source, and so they should be taken with caution. Nonetheless, from this experiment we conclude that the presence of a power-law component is statistically supported for all background regions considered. We highlight that the significance of the detection is lowest for \texttt{bkg3} ($\approx1.66\sigma$), while for the remaining regions the significance is $>2\sigma$. In particular, the significance of the detection for the backgrounds within the same chip as the source (\texttt{bkg1} and \texttt{bkg5}) is $>3.8\sigma$. At last, the flux of the power-law component fluctuates between $F_\mathrm{10-30\,keV} \sim  (1.3$--$4.8)\times 10^{-13}$~\flux, depending on the background. %, which adds further uncertainty to the source flux. 

\begin{figure}
    \centering
    \includegraphics[width=\linewidth]{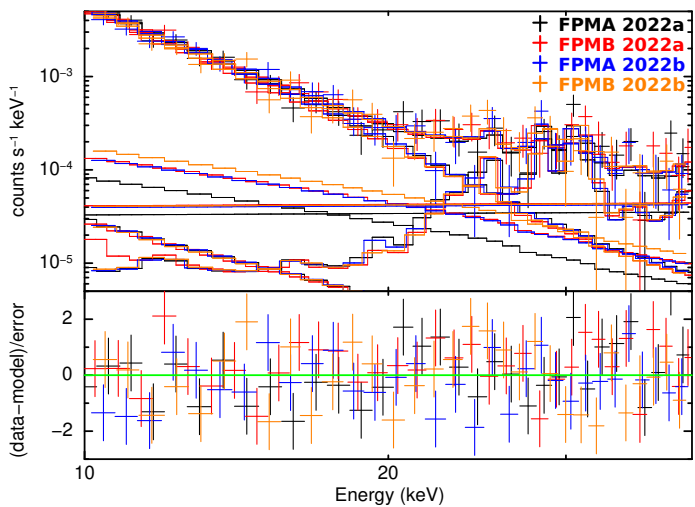}
    \caption{\Nu spectra from the region of \textit{Apep} for both observations and cameras. This is not background-subtracted, but instead a background model was fitted using \texttt{nuskybgd}; the multiple components of this background model are shown in thin lines. The source model is  \textit{TBabs*vphabs*(vpshock+po)}. }
    \label{fig:nuskybgd}
\end{figure}

Complementarily, we followed a different approach to deal with uncertainties in the background. In this case, a background model was introduced and fitted using  \texttt{nuskybgd}\footnote{\url{https://github.com/achronal/nuskybgd-py}} \citep{Wik2014}. The advantage of this approach is that the underlying background spectral model is well-known, thus limiting the uncertainties to statistical ones. The total spectrum for the source region is extracted and then fitted with a fixed background model for that location. This is repeated for each observation and camera separately, using the five background regions discussed previously to have a good sampling of the background. The result is shown in Fig.~\ref{fig:nuskybgd}. In this case the presence of a power-law component is again favoured, with a C-stat that diminishes from 682.4/589 to 666.3/588 when a power-law component is introduced. Similarly as discussed in Sect.~\ref{sec:results}, the overall decrease in C-stat is due to the better fit of the \Nu data, as the C-stat for the \XMM cameras actually increases slightly. This power-law component has a flux of $F_\mathrm{10-30\,keV} = 3.9^{+1.0}_{-1.2}\times 10^{-13}$~\flux. The 90\% confidence interval for the flux is $F_\mathrm{10-30\,keV} = (2.3$--$5.6)\times 10^{-13}$~\flux, probing consistent results with those derived from fitting the background-subtracted spectra.

%-------------------------------------------------
\section{Non-thermal emission model} \label{sec:NT_model}
%-------------------------------------------------

Here we present a review of the multi-zone model used to calculate the non-thermal radiation from \textit{Apep}. This model is suitable for this CWB as the stars are separated by tens of AU, which means that the shocks in the WCR are essentially adiabatic and quasi-stationary \citep{delPalacio2022}. The WCR structure is treated as an axi-symmetric surface under a thin shock approximation (i.e., the fluid is considered homogeneous along the direction perpendicular to the shock normal). The thermodynamical properties along the WCR (density, magnetic field intensity, etc.) are calculated with semi-analytical prescriptions based on mass and energy conservation of the fluid elements \citep{Martinez2022}, which is slightly more precise than the original assumption of Rankine-Hugoniot jump conditions in \cite{delPalacio2016}. The only free parameter in the hydrodynamical model is the ratio between the thermal pressure (set by the properties of the stellar wind) and the magnetic pressure in the WCR, $\eta_B$.

Relativistic particles accelerate when a fluid line from a stellar wind enters the WCR. The relativistic particle distribution injected at a given position in the WCR is a power law with the spectral index given by the radio observations ($p=2.4$). This distribution is normalised such that the injected power is a fraction $f_\mathrm{NT}$ of the total power available for particle acceleration ($L_\mathrm{inj,\perp}$). This power is distributed in electrons and protons as $f_\mathrm{NT} = f_\mathrm{NT,e} + f_\mathrm{NT,p}$. One common parameterisation is $f_\mathrm{NT,e} = K_\mathrm{e,p} f_\mathrm{NT}$, with $K_\mathrm{e,p} \sim 0.01 - 0.1$ \citep[see][for a discussion of uncertainties in this value]{Merten2017}. 
Upon injection, particles are attached to the fluid lines via the magnetic fields and flow together with the shocked fluid. As they stream, particles cool down due to different processes. The approximate solution used for the transport equation is given in Appendix A of \cite{delPalacio2022}.  

Finally, the non-thermal particles produce broadband radiation. The most relevant emission mechanisms are synchrotron and IC for electrons, which dominate in the radio and high-energy domain (hard X-rays and $\gamma$ rays), and proton-proton collision for protons, which can contribute to the $\gamma$-ray flux (though this hadronic contribution is subdominant if $K_\mathrm{ep} > 0.01$). This emission is then corrected for absorption, namely free-free absorption in the ionised stellar winds for low-frequency radio emission, and $\gamma$-$\gamma$ absorption in the stellar radiation fields for high-energy $\gamma$ rays.

\end{document}